\documentclass{aa}
\usepackage[varg]{txfonts}
\usepackage{graphicx}
\usepackage{natbib}
\usepackage{stfloats}
\usepackage{placeins}
\usepackage[colorlinks=true, linkcolor=blue, citecolor=blue, filecolor=blue, urlcolor=blue]{hyperref}
\bibpunct{(}{)}{;}{a}{}{,}
\usepackage{orcidlink}
\DeclareSymbolFont{upgreek}{U}{eur}{m}{n}
\DeclareMathSymbol{\umu}{0}{upgreek}{"16}
\title{Polarimetry of exoplanet-exomoon systems}
\author{M. B. Michaelis\orcidlink{0009-0007-6337-7023} \and M. Lietzow-Sinjen\orcidlink{0000-0001-9511-3371} \and S. Wolf\orcidlink{0000-0001-7841-3452}}
\institute{Institute of Theoretical Physics and Astrophysics, Kiel University, Leibnizstr. 15, 24118 Kiel, Germany\\\email{mmichaelis@astrophysik.uni-kiel.de}}
\date{Received / Accepted}
    
\abstract
    {}
    {
    We investigated the potential of polarimetric observations in the optical wavelength range for the detection of exomoons and the characterization of exoplanet-exomoon systems.}
    {Using the three-dimensional Monte Carlo radiative transfer code POLARIS, we calculated flux and polarization phase curves of Earth-like exoplanets with a satellite similar
    to Earth's moon. Of particular interest are mutual events, when one of the two bodies casts a shadow on the other or transits in front of it.}
    {We find that the signatures of mutual events in the polarization phase curve show significant variations depending on the inclination of the lunar orbit. If the planet-satellite pair is spatially resolved from the star but the satellite is spatially unresolved, the increase in the degree of polarization during a transit of the exomoon in front of the center of the exoplanet reaches $2.7\%$ in our model system near quadrature. However, the change is less than $0.5\%$ if the orbit of the exomoon is inclined such that it transits the planet noncentrally at the same phase angles. The influence of an exomoon on the polarization phase curve of an exoplanet-exomoon system is dependent on the lunar polarization phase curve. Observations of full eclipses and occultations of the exomoon allow the determination of separate polarization phase curves for the two bodies.}
    {Information about the lunar orbital inclination can be obtained with polarimetric observations of shadows or transits. Measuring the influence of large satellites not only on the total flux, but also on the polarization of the reflected stellar radiation during mutual events thus facilitates the prediction of future mutual events and the verification of exomoon candidates.
    }
    
    
\titlerunning{Polarimetry of exoplanet-exomoon systems}
\authorrunning{Michaelis et al.}

\begin{document}

\maketitle
\section{Introduction}
In the Solar System, polarimetric observations have been used for the characterization of planetary atmospheres \citep[e.g.,][]{dollfus1970,hansen1974,west1983,smith1984,tomasko1984,
braak2002,schmid2006,schmid2011,rossi2015,mc_lean2017}; circumplanetary rings \citep[e.g.,][]{dollfus1979,lane1982}; and the surfaces of atmosphere-less bodies \citep[e.g.,][]{dollfus1974,dollfus1975,kiselev2004,rosenbush2005}, for example for the surface of Earth's moon \citep[e.g.,][]{gehrels1964,coyne1970,dollfus1971,shkuratov1992a}.
Polarimetry is now becoming a valuable method for characterizing extrasolar planets as well.
With modern polarimeters such as HIPPI \citep{bailey2015}, HIPPI-2 \citep{bailey2020}, or POLISH2 \citep{wiktorowicz2023}, the polarization can be measured with a precision of a  few parts per million (ppm).
This places the detection of the polarized flux from hot Jupiters within the realm of possibility.
For instance, 
\citet{bott2018} set an upper limit of $40\ \mathrm{ppm}$ for the polarization amplitude of the reflected light signal of WASP-18b. 
\citet{bailey2021} observed the linearly polarized flux of planet-hosting systems and detected the reflected light polarization signal of 51 Pegasi b with a $2.8\sigma$ significance.

The potential of polarimetric observations of the starlight scattered by Jupiter-like exoplanets \citep[e.g.,][]{seager2000,stam2004} or Earth-like exoplanets \citep[e.g.,][]{stam2008,fauchez2017} has already been demonstrated in numerous theoretical studies using semi-analytical approaches or numerical simulations.
This includes studies on the influence of circumplanetary rings \citep{lietzow2023,veenstra2025}, different cloud compositions \citep{karalidi2011,karalidi2012,rossi2018a,lietzow2022}, and the percentage of the planet covered in clouds \citep{rossi2017, winning2024}.
The dependence on the atmospheric composition enables polarimetry to determine whether a terrestrial exoplanet has an Earth-like or Venus-like climate \citep{mahapatra2023}.
The polarized  glint of a surface ocean will also be detectable with spectropolarimetry using future instruments \citep{kopparla2018,trees2019,trees2022,vaughan2023}.

Polarimeters with even higher accuracy will be required to use exoplanet polarimetry to its full capacity.
With the Habitable Worlds Observatory (HWO), a  coronagraphic space mission planned in the 2040s, it will be possible to directly image exo\-planets in the habitable zone around their host stars and consequently to detect polarized scattered radiation from Earth-like exoplanets \citep{NAP2023,vaughan2023}.

In the Solar System, six out of eight planets
are orbited by moons. Thus, it is assumed that many exoplanets also host satellites. Despite the ongoing search for such exomoons \citep[e.g.,][]{kipping2012,teachey2018,fox2021,kipping2022,kipping2023,yahalomi2024}, there has been no confirmed detection yet. Whether the transit signals associated with the Neptune-sized exomoon candidates around the planets Kepler-1625b \citep{teachey2018} and Kepler-1708b \citep{kipping2022} exist or are merely an artifact resulting from the data reduction, remains controversial \citep{kreidberg2019,heller2024}.

The inevitable influence of an exomoon on the flux and polarization measurements of its exoplanet has been addressed previously in a few studies.
The flux change during mutual events, where one body casts a shadow on the other or transits in front of it, was calculated for the simple case of uniformly illuminated visible planetary and lunar disks by \citet{limbach2024}, for two Lambertian reflecting spheres by \citet{cabrera2007}, and for occultations in general by \citet{luger2022}.
According to \citet{limbach2024}, the HWO will be able to detect mutual events in systems analogous to the Earth and its moon up to a distance of $10\ \mathrm{pc}$.
\citet{sengupta2016} modeled the influence of exo\-moons on the polarized thermal flux of self-luminous exoplanets during transits of the satellite in front of the planet.
\citet{berzosa_molina2018}, on the other hand, used the code PyMieDAP \citep{rossi2018} to calculate the influence of an exomoon on the polarization of the radiation scattered by Earth-like exoplanets.
The authors  studied
a system with a  non-inclined exo\-moon orbit at a single wavelength of $450\ \mathrm{nm}$. However, their set of mutual events is limited to five or six phase angles per type of mutual event.

For this article we acquired selected general results for an arbitrary exoplanet-exomoon system using analytic calculations. We also performed three-dimensional Monte Carlo radiative transfer simulations to model the polarimetric signature of mutual events in systems consisting of an Earth-like exoplanet and an exomoon similar to the Earth's moon at wavelengths of $400\ \mathrm{nm}, 550\ \mathrm{nm}$, and $700\ \mathrm{nm}$ as well as for the full range of possible phase angles in the case of a non-inclined exomoon orbit. An advantage of Monte Carlo simulations is that simplifications such as the locally plane-parallel atmosphere as used in \citet{berzosa_molina2018} are not needed. We can therefore model mutual events at phase angles close to $180\degr$, which require a fully spherical description of the planetary atmosphere. 
Additionally, we investigated the influence of noncentral transits and eclipses, that is when the moon is not in the orbital plane of the planet, but is slightly above or below it during the mutual event. This corresponds to possible geometries in systems with inclined exomoon orbits. We then evaluated the potential of polarimetric observations for the characterization of the properties of the exomoon including its orbital parameters.

In Sect. \ref{sec:methods} we introduce the description of polarized radiation and our radiative transfer simulations.
Thereafter, in Sect. \ref{sec:model} we describe the parameters of the model used for this study that resembles the Earth and its moon.
In Sect. \ref{sec:results} we present the results. Section \ref{subsec:phase curve} focuses on the influence of an exomoon on the flux and polarization phase curve of an exoplanet neglecting the influence of mutual events, which  is the topic of Sect. \ref{subsec:mutual_events}.
The usability of polarimetric observations for a possible exomoon detection is discussed in Sect. \ref{sec:discussion}.
We conclude this study with a summary of our results in Sect. \ref{sec:conclusion}.

\section{Methods}
\label{sec:methods}

In this section we introduce the Stokes formalism and further concepts that are used to describe polarized radiation in this paper. We also present the framework of our simulations with the three-dimensional Monte Carlo radiative transfer code POLARIS\footnote{\url{https://portia.astrophysik.uni-kiel.de/polaris/}} \citep{reissl2016}. 
Recently, the code was extended to and optimized for the radiative transfer in exoplanetary atmospheres by \citet{lietzow2021}.
For this study, we extended the code further to include an exomoon in the model space.

The state of polarization of electromagnetic radiation is described using the Stokes vector $\vec{S} = (I, Q, U, V)^T$. Here, $I$ is the total intensity, $Q$ and $U$ describe the linearly polarized intensity and polarization direction, and $V$ is the circularly polarized intensity.
Using the entries of the Stokes vector, the degree of linear polarization $P$ and the angle of polarization $\chi$ are defined as \citep[e.g.,][]{newton1966,hansen1974b}
\begin{equation}
        P = \frac{\sqrt{Q^2 + U^2}}{I}, \quad \tan 2\chi = \frac{U}{Q}.
\end{equation}
Throughout this study, the contribution of the stellar flux and thus its impact on the net polarization of a spatially unresolved exoplanetary system is not considered. However, the requirements for polarimetric observations with the star included are briefly discussed in Sect. \ref{sec:discussion}.

POLARIS solves the radiative transfer equation by dividing the radiative field into photon packages which are characterized by their Stokes vector and move through the model space on a random path \citep{reissl2016}. 
They are emitted at a random place on the surface of the star.
The radiation emitted by the star is assumed to be unpolarized, 
as a disk-integrated polarization of less than $1\ \mathrm{ppm}$ was measured for the quiet sun \citep{kemp1987}. Radiation emitted by other inactive FGK-dwarfs also has an intrinsic degree of linear polarization around only $10\ \mathrm{ppm}$ \citep{cotton2017}.
Because the exoplanet (in short "planet") and its exomoon (in short "moon") are seen under a very small solid angle from a point on the stellar surface ($5.7\cdot 10^{-9}\ \mathrm{sr}$ for the Earth seen from the sun),
photons are only emitted toward the planet and its atmosphere as described in \citet{lietzow2021} or toward the surface of the moon.

The modification of the polarization state and thus the values of the Stokes parameters of a photon package due to scattering by gas particles in the exoplanetary atmosphere is specified by the Müller matrix $\mathbf{F}(\Theta)$. The Stokes vector $\vec{S}^\prime$ after scattering is \citep[e.g.,][]{newton1966,hansen1974b}
\begin{equation}
        \vec{S}^\prime \propto \mathbf{F}(\Theta) \cdot \vec{S},
        \label{eq:scat_matrix}
\end{equation} 
where $\Theta$ is the scattering angle and $\vec{S}$ is the Stokes vector of the incoming radiation.
The influence of absorption by atmospheric particles 
is ignored in this study.
For the atmosphere of Earth, absorption is of minor importance at wavelengths of $400\ \mathrm{nm}$ and $550\,\mathrm{nm}$, while absorption by water vapor and $\mathrm{O_2}$ has a noticeable influence on the flux at $700\ \mathrm{nm}$ \citep[see, e.g.,][]{schubert2000}. Neglecting this influence has the advantage that the wavelength dependence of our results for an Earth-like exo\-planet can also be interpreted as a dependence on the optical depth and hence on the density of the planetary atmosphere.

If a photon package is reflected by the surface of the planet or moon, its energy is weighted by
the surface albedo $\omega$ which is the fraction of incoming radiation that is scattered by a surface rather than absorbed \citep[e.g.,][]{nicodemus1977}.
The corresponding impact on the Stokes vector is described by a reflection matrix \citep[see, e.g.,][]{zhai2010}.

\section{Model of planet and moon}
\label{sec:model}

\begin{table}
        \caption{Parameters of the star, the planet, and the moon of the baseline model used for our simulations.}
        \label{tab:parameters}
        \centering
    \begin{tabular}{rl}
                \hline\hline\\[-1em]
        \multicolumn{2}{c}{Stellar parameters} \\
        \\[-1em]\hline\\[-1em]
        Radius $r_\mathrm{s}\ [\mathrm{R}_\odot]$ & $1$ \\
        Eff. temperature $T_\mathrm{s}\ [\mathrm{K}]$ & $5772$ \\
        \hline\\[-1em]
        \multicolumn{2}{c}{Planetary parameters} \\
        \\[-1em]\hline\\[-1em]
        Radius $r_\mathrm{p}\ [\mathrm{km}]$ & $6371.0$ \\
        Surface albedo $\omega_\mathrm{p}$ & $0.3$ \\
        Orbital radius $d_\mathrm{sp}\ [\mathrm{AU}]$ & $1$ \\
        Eccentricity $e_\mathrm{b}$ & $0$ \\
        Inclination $i_\mathrm{b}\ [\degr]$ & $90$ \\
        \hline\\[-1em]
        \multicolumn{2}{c}{Atmospheric parameters (planet)} \\
        \\[-1em]\hline\\[-1em]
        $A\ [\text{units of }10^{-5}]$ & $28.71$ \citep{keady2000} \\
        $B\ [\text{units of }10^{-5}]$ & $5.67$ \citep{keady2000} \\
        Depolarization factor $\delta$ & $0.028$ \citep{bates1984} \\
        Optical depth $\tau$ ($550\,\mathrm{nm}$) & $0.098$ \\
        \hline\\[-1em]
        \multicolumn{2}{c}{Lunar parameters} \\
        \\[-1em]\hline\\[-1em]
        Radius $r_\mathrm{m}\ [\mathrm{km}]$ & $1737.1$ \\
        Surface albedo $\omega_\mathrm{m}$ & $0.1$ \\
        Orbital radius $d_\mathrm{pm}\ [\mathrm{km}]$ & $380\,000$ \\
        Eccentricity $e_\mathrm{m}$ & $0$ \\
        Inclination $i_\mathrm{m}\ [\degr]$ & $0$\\
        \hline
    \end{tabular}
    \tablefoot{Constants $A$ and $B$ are used to calculate the refractive index of air according to Eq. \eqref{eq:n}.
    }
\end{table}

For our simulations, we chose a model consisting of an Earth-like exo\-planet and an exomoon comparable to the moon of Earth. The reasoning behind this choice is that such a system has a much more favorable moon-to-planet continuum flux ratio of $F_\mathrm{m}/F_\mathrm{p} \approx 2\%$ than the moons around the gas planets in the outer Solar System,  which reach 
$F_\mathrm{m}/F_\mathrm{p}$ of at most $10^{-3}$, with the only exception of the Neptune-Triton pair with $0.5\%$ 
(see Appendix \ref{app:flux_ratios}). 
The parameters of the model are listed in Table \ref{tab:parameters}. The planetary parameters were chosen to be comparable to the model used by \citet{stam2008} and \citet{berzosa_molina2018}.

\begin{figure}
\centering
        \includegraphics{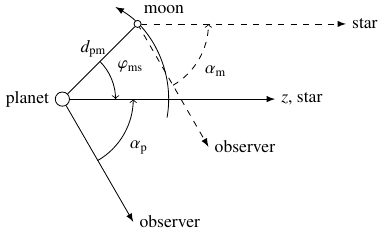}
        \caption{Geometry of the non-inclined baseline system. The centers of the star, the planet, and the moon as well as the observer are all coplanar. The angles $\alpha_\mathrm{p}$ and $\alpha_\mathrm{m}$ are the phase angles of the planet and the moon, respectively. 
    $\varphi_\mathrm{ms}$ is the angle between the direction to the star and the direction to the moon for an observer at the center of the planet.}
        \label{fig:geometry}
\end{figure}

The planetary orbit around the star in the model is circular ($e_\mathrm{b} = 0$) with a radius of $d_\mathrm{sp} = 1\ \mathrm{AU}$. It is seen edge-on ($i_\mathrm{b} = 90^\circ$). We neglect the distance between the planet-moon barycenter and the geometric center of the planet. The orbit of the moon is also circular, with a radius of $d_\mathrm{pm} = 380\,000\ \mathrm{km}$. In the baseline model, it is not inclined with respect to the orbital plane of the planet. The influence of an inclined orbit of the moon is topic of Sect. \ref{subsubsec:non_central}.
The position of the planet with respect to the star and the observer is
defined using the planetary phase angle $\alpha_\mathrm{p}$. The angle $\varphi_\mathrm{ms}$ specifies the position of the moon on its orbit around the planet as defined in Fig.~\ref{fig:geometry}. The moon orbits the planet prograde with respect to the planetary orbit.

The model atmosphere of the planet in our simulations consists solely of gas particles and does not include a cloud layer.
Because gas particles are small compared to the considered wavelengths ($400\ \mathrm{nm}$, $550\ \mathrm{nm}$, $700\ \mathrm{nm}$), we describe their scattering properties by applying the concept of Rayleigh scattering. The scattering matrix $\mathbf{F}(\Theta)$ for Rayleigh scattering can be found in \citet{hansen1974b}.
The Rayleigh scattering cross section for anisotropic, randomly oriented gas particles at wavelength $\lambda$ is given by \citet{sneep2005}.

We use typical values for air and assume that the depolarization factor $\delta = 0.028$ \citep{bates1984} is constant across all considered wavelengths. The refractive index $n(\lambda)$ is calculated according to \citet{keady2000} as
\begin{equation}
        n(\lambda) - 1 = A\left(1+\frac{B}{(\lambda/\umu\text{m})^2}\right),
        \label{eq:n}
\end{equation}
in which for air $A = 28.71\cdot 10^{-5}$ and $B = 5.67\cdot 10^{-5}$.
The model atmosphere consists of 32 vertical layers with a total height of $100\ \mathrm{km}$. We use the pressure profile  
for the mid-latitude summer profile of the atmosphere of Earth as given in \citet{mc_clatchey1972} and calculate the number density 
in each layer 
assuming hydrostatic equilibrium \citep[similar to][]{berzosa_molina2018,lietzow2021} 
and a molar mass of air of $28.964\cdot 10^{-3}\,\mathrm{kg\ mol^{-1}}$ \citep{schubert2000}.
The optical depth of the model atmosphere is $\tau_\lambda = 0.098$ at a wavelength of $550\ \mathrm{nm}$, identical to the result by \citet{stam2008}. 

Both bodies have diffuse reflecting (Lambertian) surfaces with albedoes of $\omega_\mathrm{p} = 0.3$ and $\omega_\mathrm{m} = 0.1$ for the planet and the moon, respectively.
After a reflection by such a surface, the radiation is completely depolarized \citep[see also][]{lietzow2021}.
Diffuse reflecting surfaces are commonly used as a model for exoplanetary surfaces \citep[e.g.,][]{stam2008,rossi2017,berzosa_molina2018}. However, they are only an approximation of real planetary surfaces. Even radiation reflected by atmosphere-less bodies such as Mercury or Earth's moon reaches degrees of linear polarization of a few percent at wavelengths in the visual spectrum \citep[e.g.,][]{lyot1929,coyne1970,dollfus1974}.

Because the moon in the model is a diffuse reflecting sphere, its flux phase curve is calculated analytically according to \citet{russell1916} in the following sections.
Because it does not have an atmosphere, it does not contribute to the polarized flux. In our study, we evaluate in which cases this assumption is justified and present analytical estimates of the systematical errors.

\section{Results}
\label{sec:results}

In the following presentations of our modeling results, we normalize the flux of the planet $F_\mathrm{p}$ and the moon $F_\mathrm{m}$ such that they are equal to the geometrical albedoes $A_\mathrm{p}$ and $A_\mathrm{m}$ at a phase angle of $0\degr$. 
For the planet, this is achieved by dividing the planetary flux phase curve $F_\mathrm{p}(\alpha_\mathrm{p})$ by $F_\mathrm{0p}$, the flux reflected by a flat fully diffuse reflecting disk with the same radius as the planet.
If $B_\nu$ is the Planck function, $\nu$ the frequency, $T_\mathrm{s}$ the stellar effective temperature, $r_\mathrm{s}$ the stellar radius, and $d_\mathrm{po}$ the distance between the planet and the observer, then
\begin{equation}
        F_{0\mathrm{p}} = \pi B_\nu(T_\mathrm{s}) \frac{r_\mathrm{s}^2 r_\mathrm{p}^2}{d_\mathrm{sp}^2 d_\mathrm{po}^2}.
        \label{eq:F0p}
\end{equation}
This is a common approach to make the results independent of the basic stellar and planetary parameters \citep[e.g.,][]{stam2008,berzosa_molina2018,lietzow2021}. 
For the exomoon $F_{0\mathrm{m}}$ is defined accordingly.
The flux of the unresolved planet-moon system is normalized by dividing by $F_{0\mathrm{p}}$ which facilitates the comparability with the results for the flux of the planet alone.

\subsection{Phase curves}
\label{subsec:phase curve}

\begin{figure}
        \includegraphics{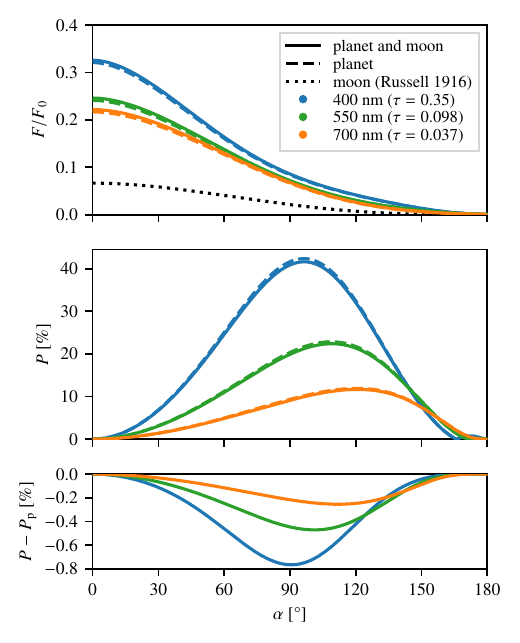}
        \caption{Phase curves of the planet, the moon, and the planet and moon together but without the influence of mutual events. {(Top)} Flux $F/F_0$ as a function of the phase angle $\alpha$. For the planet and for planet and moon combined, $F_0 = F_\mathrm{0p}$. In contrast, for the moon, $F_0 = F_\mathrm{0m}$. In addition, we use the analytic result from \citet{russell1916} for a Lambertian sphere for the phase curve of the moon. {(Middle)} Phase curves for the linear polarization $P(\alpha)$ for the planet and planet and moon combined, assuming $P_\mathrm{m} = 0$ (i.e., of the diffuse reflecting moon surface). The stellar flux is not included in the calculation of $P$. {(Bottom)} Difference $P-P_\mathrm{p}$ between the polarization phase curve of the unresolved system consisting of planet and moon and the planetary polarization phase curve.
        }
        \label{fig:general_phasecurves}
\end{figure}

First, we calculated phase curves of the flux and degree of linear polarization of the moon, the planet, and the two bodies combined neglecting the influence of possible eclipses, shadows, transits, or occultations which are discussed in detail in Sect.~\ref{subsec:mutual_events}.
Solely for the curves presented in this subsection, the Stokes vector of the planet $\vec{S}_\mathrm{p}$ and the moon $\vec{S}_\mathrm{m}$ were calculated separately and added to get the Stokes vector of the unresolved system.
The systematic error caused by neglecting the flux that an observer measures due to the mutual illumination of the planet and the moon is less than $2\cdot 10^{-4}$ of the maximum lunar flux for the model system (see Appendix \ref{app:mutual_illumination}). Also, we neglected the small difference in the planetary and the lunar phase angle and assumed $\alpha_\mathrm{p} \approx \alpha_\mathrm{m} \approx \alpha$.
For the planetary Stokes vector, we used results of POLARIS simulations.
The disk-integrated results of our simulations are shown in Fig.~\ref{fig:general_phasecurves}.

The linear polarization phase curve of the planet resembles a typical curve for planets with Rayleigh scattering atmospheres, which is discussed in detail by \citet{stam2008}.
The maximum degree of polarization $P$  of $42.4\%$ at a wavelength of $400\ \mathrm{nm}$ is reached at a phase angle of $96^\circ$. For the longer wavelengths, the maximum shifts to larger phase angles of $109^\circ$ for $550\ \mathrm{nm}$, where $P$ reaches $22.9\%$, and $120^\circ$ for $700\ \mathrm{nm}$, where $P$ is $11.9\%$.

The direction of polarization is mostly perpendicular to the orbital and scattering plane of the planet. Only at very large phase angles, it is parallel to the scattering plane. This results in a zero-crossing of $P(\alpha)$ when the direction of polarization changes. For $400\ \mathrm{nm}$ this change occurs at $166^\circ$ and for $550\ \mathrm{nm}$ at $171^\circ$. According to \citet{stam2008}, this change is due to a greater influence of multiple scattered photons. For $700\ \mathrm{nm}$ the zero-crossing and the change in the polarization direction vanish, as the probability of multiple scattering is too low due to the small atmospheric optical depth of only $0.037$.

The moon contributes to the total flux but not to the polarized flux of the system. Therefore, the degree of linear polarization is slightly lower at all phase angles and wavelengths for the planet-moon system compared to the planet, similar to the results by \citet{berzosa_molina2018} at $450\ \mathrm{nm}$.
The maximum reduction for $400\ \mathrm{nm}$ is found at a phase angle of $91\degr$ with a difference of $-0.77\%$. For longer wavelengths, the maximum reduction shifts toward larger phase angles, similar to the maximum of $P(\alpha)$. For $550\ \mathrm{nm}$ the maximum reduction is found at a phase angle of $101\degr$ with a difference of $-0.46\%$, for $700\ \mathrm{nm}$ it is at $\alpha = 112\degr$ with $-0.25\%$. The phase angle dependent influence is also illustrated in the bottom panel of Fig.~\ref{fig:general_phasecurves}.

We find that the behavior might be converse for a real exomoon though. Let $F_\mathrm{p}, P_\mathrm{p}$ be the flux and degree of polarization of the planet, respectively, and $F_\mathrm{m}, P_\mathrm{m}$ the same quantities for the moon. 
If the planet and the moon are symmetric to the scattering plane, the direction of polarization is either perpendicular or parallel to the scattering plane. Therefore, the polarized fluxes $F_\mathrm{p}P_\mathrm{p}$ and $F_\mathrm{m}P_\mathrm{m}$ either have to be added or subtracted to get the polarized flux of the unresolved system. 
Because the planetary flux $F_\mathrm{p}$ is much bigger than the flux of the moon $F_\mathrm{m}$, we assume that $F_\mathrm{p}P_\mathrm{p} \geq F_\mathrm{m}P_\mathrm{m}$. For the model system, this assumption is fulfilled everywhere as $P_\mathrm{m} = 0$.
The difference between the degree of linear polarization of the unresolved system and the planet alone is therefore
\begin{equation}
        P - P_\mathrm{p}
        = \frac{F_\mathrm{p}P_\mathrm{p} \pm F_\mathrm{m}P_\mathrm{m}}{F_\mathrm{p} + F_\mathrm{m}} - P_\mathrm{p}
        = -\frac{F_\mathrm{m}}{F_\mathrm{p} + F_\mathrm{m}}(P_\mathrm{p} \mp P_\mathrm{m}).
        \label{eq:p_change}
\end{equation}
Here, the minus-plus-symbol ($\mp$) in the last expression differentiates between cases in which the direction of polarization is identical for the planet and moon ($-$) and cases in which they are perpendicular (+). If $P_\mathrm{m} > P_\mathrm{p}$ while the directions of polarization are parallel, the quantity $P - P_\mathrm{p}$ is therefore positive.

The direction of the polarization of Earth's moon is parallel to the scattering plane for phase angles of less than the inversion phase angle $\alpha_\mathrm{inv}$, which is between $\alpha = 20\degr$ and $30\degr$, and perpendicular for phase angles larger than $\alpha_\mathrm{inv}$
\citep[e.g.,][]{lyot1929,coyne1970,shkuratov1992a}. 
If $P_\mathrm{m} > 0$ is assumed, we thus expect a larger reduction of $P$ than shown in Fig.~\ref{fig:general_phasecurves} for phase angles below $\alpha_\mathrm{inv}$ and at large phase angles where the planetary direction of polarization is parallel to the scattering plane. Near quadrature, where we found the maximum differences $P - P_\mathrm{p}$ due to a moon, the degree of linear polarization is less affected. Assuming a realistic value of $P_\mathrm{m} \approx 10\%$ \citep{lyot1929,coyne1970}, the maximum value of $P - P_\mathrm{p}$ at $550\ \mathrm{nm}$ is only $-0.25\%$ rather than $-0.46\%$ at $101\degr$.

Even if it were possible to measure flux and polarization with a precision high enough to detect the subtle differences between the curves of the planet alone and of the planet-moon system, the following ambiguity would remain. A slightly higher flux and a slightly altered degree of linear polarization can also be explained by an adapted planetary model, for instance by a slightly higher surface albedo. Therefore, the presence of an exomoon cannot be deduced from the phase curves shown in Fig.~\ref{fig:general_phasecurves} \citep[see also][]{berzosa_molina2018}. This ambiguity can be resolved by observing the influences of mutual events, which are investigated in the next section.

\subsection{Mutual events}
\label{subsec:mutual_events}

\begin{figure}
        \centering
        \includegraphics{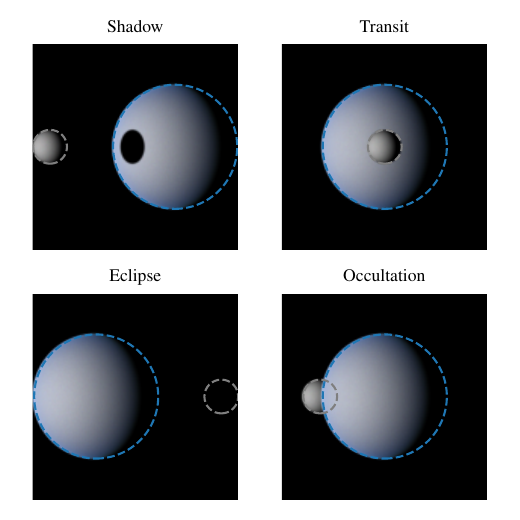}
        \caption{Four types of mutual events simulated for a phase angle of $225\degr$. The RGB images are based on individual maps simulated for wavelengths of $440\ \mathrm{nm}$ (blue), $555\ \mathrm{nm}$ (green), and $700\ \mathrm{nm}$ (red). The total flux was normalized to the geometric albedo separately at each of the wavelengths. The dashed circles indicate the position and size of the planet (blue) and moon (gray). To improve the visibility of the moon and to fit both bodies into the images for shadow and eclipse, the lunar surface albedo $a_\mathrm{m}$ was increased to $0.3$ and the moon's semimajor axis was set to $2.8\,r_\mathrm{p}$ for this graphic.}
        \label{fig:mutual_events}
\end{figure}

The four types of mutual events that can occur in a system are shown in Fig.~\ref{fig:mutual_events}. We stick to the terminology introduced by \citet{cabrera2007} and \citet{limbach2024}:
\begin{enumerate}
        \item The moon passes between the planet and the star. It casts a shadow on the planet.
        \item The moon passes between the planet and the observer, which is called a transit of the moon in front of the planet.
        \item The moon is in the shadow of the planet, a lunar eclipse.
        \item The planet passes between the moon and the observer. This is called an occultation of the moon by the planet.
\end{enumerate}
We first calculate the influence of full occultations of an exomoon and total lunar eclipses in Sect. \ref{subsubsec:occultation_eclipse} analytically. Thereafter, the influence of transits and shadows is determined for the model system described in Sect. \ref{sec:model} using POLARIS simulations in Sect. \ref{subsubsec:central_transit} and \ref{subsubsec:central_shadow}. Finally, the differences for noncentral transits and shadows are highlighted in Sect. \ref{subsubsec:non_central}. We are interested in the change in the total flux $\Delta F$ and the linear polarization $\Delta P$, which are
\begin{equation}
        \Delta F = F_\text{with} - F, \quad \Delta P = P_\text{with} - P,
\end{equation}
where $F_\text{with}$ and $P_\text{with}$ are respectively the flux and linear polarization observed during a mutual event, and $F$ and $P$ are the same quantities that would be observed at the same phase angle without the mutual event.

\subsubsection{Full occultations and eclipses}
\label{subsubsec:occultation_eclipse}

If the moon is completely occulted by the planet or completely within the umbra of the planet, the flux change is $\Delta F = -F_\mathrm{m}$, where $F_\mathrm{m}$ is the lunar flux that would be measured at the respective phase angle if there was not an eclipse or occultation.
Here, we neglected any radiation that is scattered toward the moon in the planetary atmosphere during a lunar eclipse \citep{cabrera2007,limbach2024}.

Because the influence of the moon vanishes during these two mutual events, the change in the degree of polarization $\Delta P$ during the event is the inversion of the influence described in Eq. \eqref{eq:p_change}.
For the model system, the maximum change during these events is therefore extracted from the bottom panel of Fig.~\ref{fig:general_phasecurves} by flipping the algebraic sign. We find the maximum changes to be $\Delta P = +0.77\%$ for $400\ \mathrm{nm}$ if a mutual event happens at $\alpha = 91\degr$, $\Delta P = +0.46\%$ for $550\ \mathrm{nm}$ at $\alpha = 101\degr$, and $\Delta P = +0.25\%$ for $700\ \mathrm{nm}$ at $\alpha = 112\degr$.

Assuming reflection properties similar to the Earth's moon, the same influence that was described in the previous section is expected. Assuming $P_\mathrm{m} \approx 10\%$ around quadrature, the maximum $\Delta P$ at $550\ \mathrm{nm}$ is only $+0.25\%$.
The systematic error caused by the assumption of $P_\mathrm{m} = 0$ is on the same order of magnitude as the actual result and $\Delta P$ might even be negative if $P_\mathrm{m} > P_\mathrm{p}$.

It should be noted that we calculated much smaller values for $\Delta P$ than reported by \cite{berzosa_molina2018} who found maximum values for $\Delta P$ of $+2.5\%$ during full occultations of the moon and of $+2.7\%$ during lunar eclipses.
However, using values for $P_\mathrm{p}$, $P_\mathrm{m}$, $\Delta F$, and $F_\mathrm{p}$ given in \citet{berzosa_molina2018} and Eq. \eqref{eq:p_change} results in a realistic $\Delta P$ of approximately $0.7\%$ to $0.8\%$, which is similar to the value we determined for $400\ \mathrm{nm}$.

For the lunar darkening events, $\Delta P$ is sensitive to the surface albedo of the moon.
If $P_\mathrm{m} = 0$ and the albedo of the moon is $\omega_\mathrm{m} > 0.2$, the changes $\Delta P$
reach values of $3\%$ and more.

According to Eq. \eqref{eq:p_change}, observations during an eclipse or occultation allow for the determination of the difference or sum of the degrees of polarization $P_\mathrm{p} \mp P_\mathrm{m}$, because $F_\mathrm{m} = \Delta F$ and $F_\mathrm{p} = F_\mathrm{with}$ can be measured directly. 
$P_\mathrm{p}$ is also measured directly during a transit because the degree of polarization of the unresolved system is $P = P_\mathrm{p}$ when the moon is completely eclipsed or occulted. Therefore, the sign $\mp$ is also determined.
As a consequence, both $P_\mathrm{p}$ and $P_\mathrm{m}$ can be determined at the phase angle where the eclipse or occultation is observed.
In other words, such observations allow us to disentangle the planetary and the lunar polarization phase curve.

\subsubsection{Central transits}
\label{subsubsec:central_transit}

\begin{figure*}
        \centering
        \includegraphics{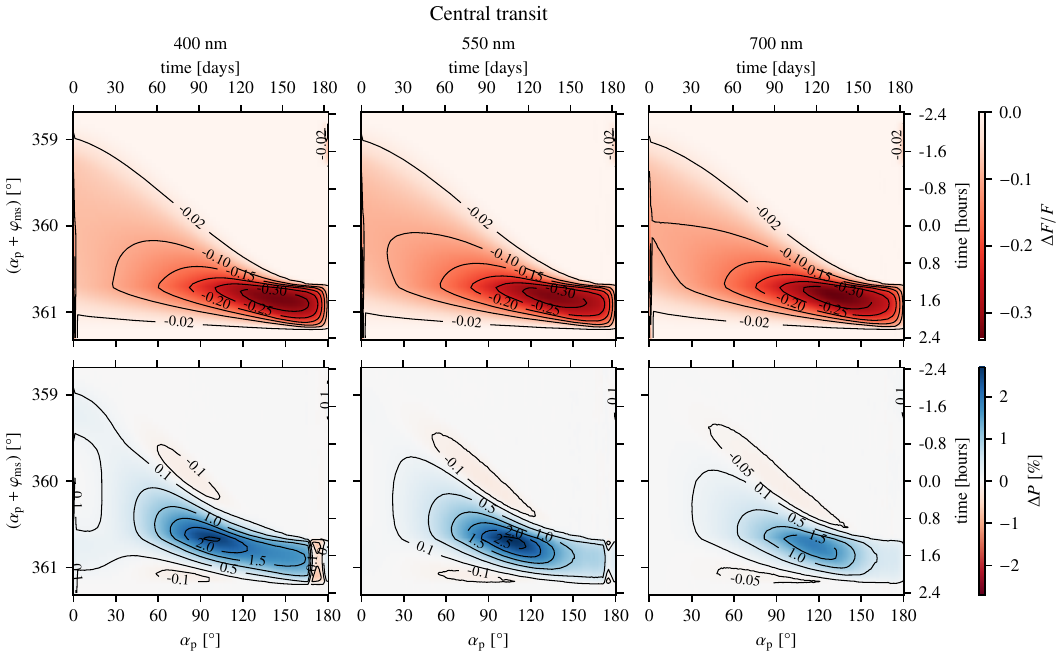}
        \caption{Relative change of the total flux $\Delta F/F$ (top row) and absolute change of the degree of polarization $\Delta P$ (bottom row) during central transits of the moon in front of the planet at different phase angles $\alpha_\mathrm{p}$ visualized using color maps and contour lines. The upper abscissa shows the time in days since the planet was at $\alpha = 0\degr$. On the left ordinate, the relevant angle $\alpha_\mathrm{p} + \varphi_\mathrm{ms}$ is shown. The angle was transformed to a time in hours on the right ordinate, where a time of $0$ corresponds to the time when the planet, moon, and observer are perfectly aligned. A single transit of the moon corresponds to a passage through the color maps perpendicular to the abscissa. For phase angles above $180\degr$, $F(360\degr-\alpha_\mathrm{p}, 360\degr - \varphi_\mathrm{ms}) = F(\alpha_\mathrm{p},\varphi_\mathrm{ms})$.}
        \label{fig:transit}
\end{figure*}
\begin{figure*}
        \centering
        \includegraphics{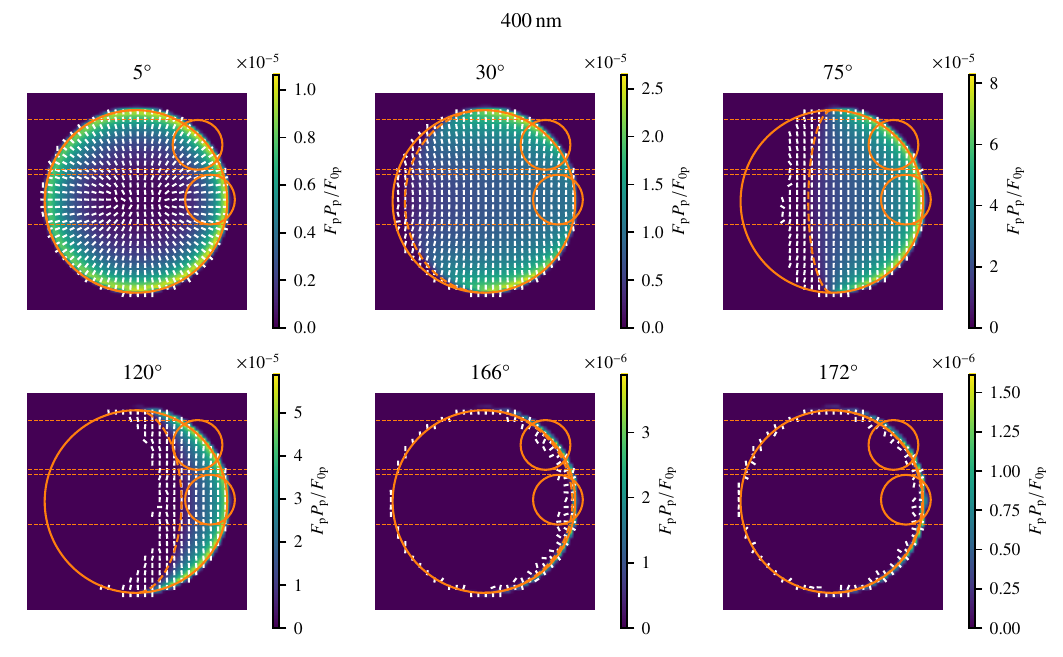}
        \caption{Polarization maps for different phase angles at a wavelength of $400\ \mathrm{nm}$. The polarized flux $F_\mathrm{p} P_\mathrm{p}$ is encoded as a color map. The resolution is $60\times 60$ px, the planet has a diameter of $50$ px. Polarization directions are shown as white lines. The larger orange circle marks the planet's circumference. The orange dashed lines between the poles of the planet mark the terminator. The day side of the planet is to the right of it. The horizontal orange lines mark the path of the moon during a central transit (lower pair of lines), or a noncentral transit (upper pair of lines). For two positions on the edge of the planet, the circumference of the moon is shown as an orange circle.}
        \label{fig:transit_polarization}
\end{figure*}

We now continue with a more in-depth description of mutual events which influence the planetary flux.
In this section we start with central transits of the moon in front of the planet.

In the model system, such a central transit
occurs
for angles $\alpha_\mathrm{p} + \varphi_\mathrm{ms}$ in the interval $[360\degr - \Delta\varphi_\mathrm{ms}, 360\degr + \Delta\varphi_\mathrm{ms}]$.
For the model system, the half width $\Delta\varphi_\mathrm{ms}$ is $1.23\degr$ (see Appendix \ref{app:angles}).
A transit in the model system therefore lasts no longer than $4.4$ hours. 
With increasing phase angle, a decreasing part of the day side of the planet is visible to the observer.
The time during which a transit causes a change in flux or linear polarization is therefore decreasing with increasing phase angle \citep[for an analytical approach, see][]{cabrera2007,limbach2024}.

The values of $\Delta F/F$ and $\Delta P$ resulting from our simulations are visualized in Fig.~\ref{fig:transit}. As $\Delta F$ is strongly dependent on the reflected flux phase curve of the planet as shown by \citet{berzosa_molina2018}, the quantity $\Delta F/F$ during a central transit is more insightful. The color maps have a resolution of $181$ phase angles $\alpha_\mathrm{p}$ between $0\degr$ and $180\degr$ 
and a resolution of $58$ lunar position angles $\varphi_\mathrm{ms}$
for each $\alpha_\mathrm{p}$. The values for $\Delta F/F$ and $\Delta P$ for a single transit are found along a passage perpendicular to the abscissa through the respective diagrams as the planetary phase angle only changes about $0.2\degr$ during a transit. For the model system, transits occur 6 or 7 times during half the revolution of the planet around the star.

For phase angles below $2\degr$, we find an obvious feature in the $\Delta F/F$ plots. If a transit happens at these phase angles, the shadow of the moon also falls on the planet.
Disregarding this special configuration with a doubled influence, 
the maximum changes $\Delta F/F$ for phase angles up to $30\degr$ are between $-8\%$ and $-12\%$.
With increasing phase angle, central transits cause larger reductions in $\Delta F / F$. 
During a transit at higher phase angles, the moon covers a bigger fraction of the illuminated part of the planetary disk 
that 
is now restricted to a small crescent.
At a wavelength of $400\ \mathrm{nm}$, $\Delta F/F$ reaches the biggest decrease of $-32.8\%$ for a transit at $\alpha_\mathrm{p} = 144\degr$.
This maximum shifts to smaller phase angles for the other two wavelengths, reaching values of $-32.3\%$ at $132\degr$ at $550\ \mathrm{nm}$ and $-33.9\%$ at $131\degr$ at $700\ \mathrm{nm}$.

The similarities between the different wavelengths are to be expected, given that the flux is mainly influenced by geometric effects, for instance, what fraction of the planetary day side is occulted.
The higher decreases of $\Delta F/F$ during a transit toward longer wavelengths at phase angles below $60\degr$ are caused by the decrease in optical depth.
The planetary brightness distribution approaches that of a diffuse reflecting sphere, so the flux is more concentrated toward the sub-stellar point on the planetary surface.
The moon thus blocks a higher fraction of the planetary flux when it occults the sub-stellar point.

At large phase angles above $160\degr$, a light ring forms around the planet as a consequence of multiple and forward scattering in the planetary atmosphere, an additional contribution to $F$, that decreases the influence $\Delta F/F$ of the lunar transit. 
Also, a flux change between $-2\%$ and $-7\%$ occurs before the moon transits the very small crescent on the day side of the planet starting at a phase angle of $176\degr$. During this phase, the moon is in front of the bright atmosphere on the side opposite the crescent. We note that this light ring only occurs in a simulation when a fully spherical model of the planetary atmosphere is used.

The absolute change $\Delta P$ in the degree of linear polarization during central transits inhabits a slightly higher dependence on the optical depth and hence on the wavelength. Furthermore, the dependence on the planetary phase angle $\alpha_\mathrm{p}$ is more complicated. Multiple different regions are found and decreases as well as increases in $P$ are possible.

In Fig.~\ref{fig:transit_polarization}, the polarized flux $F_\mathrm{p}P_\mathrm{p}$ and the direction of polarization $\chi$ are shown for the resolved planet using results from the simulations in Sect. \ref{subsec:phase curve} for a wavelength of $400\ \mathrm{nm}$. Similar figures for $550\ \mathrm{nm}$ and $700\ \mathrm{nm}$ can be found in Appendix \ref{app:disk_resolved_F_P}. Figure~\ref{fig:transit_polarization} is used to facilitate the following discussion.

At a wavelength of $400\ \mathrm{nm}$, a slight increase of $\Delta P \approx 0.2\%$ is observed at phase angles up to $20\degr$ when the moon is in front of the edge of the planet as seen from the observer.
In between, $P$ returns to values close to those calculated without a mutual event.
From the panel for a phase angle of $5\degr$ in the first row of Fig.~\ref{fig:transit_polarization}, it is clear that the polarization directions
form an almost symmetrical pattern.
Consequently, $P_\mathrm{p}(5\degr)$ is small at only $0.2\%$ at a wavelength of $400\ \mathrm{nm}$ and even lower for $550\ \mathrm{nm}$ ($0.1\%$) and $700\ \mathrm{nm}$ ($0.04\%$).
As described by \citet{berzosa_molina2018}, the moon breaks this symmetry when it crosses the limb of the planet as sketched in Fig.~\ref{fig:transit_polarization}. 

We find that the described feature is too faint at $550\ \mathrm{nm}$ and $700\ \mathrm{nm}$ to appear in the maps for $\Delta P$ in Fig.~\ref{fig:transit}. For phase angles below $10\degr$, $\Delta P$ is below $0.1\%$ because the polarized flux is too small even if the symmetry is broken. For larger phase angles, due to the smaller influence of multiple scattering at lower optical depths, the direction of polarization is 
perpendicular to the scattering plane across the planetary disk at these wavelengths which causes a different behavior.

At a wavelength of $400\ \mathrm{nm}$, a direction of polarization vertical to the scattering plane throughout the planetary disk is observed for phase angles above $20\degr$.
In Fig.~\ref{fig:transit_polarization}, this can be seen in the surface brightness maps of the disk-resolved polarized flux of the planet at phase angles of $30\degr, 75\degr$, and $120\degr$.
The moon now causes a central peak in the degree of polarization.
In addition, a slight decrease in $P$ is observed at the beginning and the end of a transit at phase angles between $50\degr$ and $120\degr$. These features are also present in the phase curves in \citet{berzosa_molina2018}, who find maximum absolute values at $450\ \mathrm{nm}$ that are similar to our results at $400\ \mathrm{nm}$ and $550\ \mathrm{nm}$.

Because the polarization directions are parallel, the change $\Delta P$
depends solely on the fraction of blocked polarized flux compared to the fraction of blocked total flux.
The central peak in $\Delta P$ during the transit occurs when the moon is in between the terminator and the edge of the planet. Here, it cannot block a high fraction of $F_\mathrm{p}P_\mathrm{p}$.
Due to the higher fraction of $F$ being blocked, $P$ increases.
At $400\ \mathrm{nm}$, the largest change is an increase of $\Delta P \approx 2.6\%$ around phase angles of $\alpha_\mathrm{p} \approx 98\degr$.
With increasing wavelength, thus decreasing optical depth, we find that the maximum value in $\Delta P$ shifts toward larger phase angles reaching $2.7\%$ during a transit at a planetary phase angle of $102\degr$ for a wavelength of $550\ \mathrm{nm}$ and $1.9\%$ for a planetary phase angle of $116\degr$ at $700\ \mathrm{nm}$.
This is similar to the maximum of the planetary polarization curve that shifts toward larger phase angles with increasing wavelength as well (see Fig.~\ref{fig:general_phasecurves}). 

Interestingly, the change in polarization during transits of the moon in front of the planet is relatively robust with respect to the actual lunar polarization phase curve (see Appendix \ref{app:lunar_influence}). Although the assumption that $P_\mathrm{m} = 0$ may result in a slight underestimation of positive changes $\Delta P$ and a slight overestimation of negative changes $\Delta P$, it does not change the general behavior shown in Fig.~\ref{fig:transit}. Using a realistic maximum value of $P_\mathrm{m} = 10\%$ \citep{lyot1929,coyne1970}, the expected absolute systematic errors $\Delta P_\mathrm{sys}$ for the maximum changes $\Delta P$ given above are approximately $\Delta P_\mathrm{sys} \approx 0.1\%$ for all wavelengths.

At $\alpha_\mathrm{p} = 75\degr$, $F_\mathrm{p}P_\mathrm{p}$ is 
highest at the edge of the visible planetary disk. When the moon transits across this edge, $P$ decreases because a higher fraction of $F_\mathrm{p}P_\mathrm{p}$ than of $F$ is blocked by the moon. 
Near the terminator, $F_\mathrm{p}P_\mathrm{p}$ and $F$ are both small. 
However, again a higher fraction of $F_\mathrm{p}P_\mathrm{p}$ than of $F$ is blocked and the degree of polarization decreases \citep[for an explanation using the Stokes parameters, see][]{berzosa_molina2018}. The biggest decrease is higher than $\Delta P = -0.1\%$ at all wavelengths but has a slightly smaller absolute value than $-0.1\%$ at the end of transits at $700\ \mathrm{nm}$ as $P$ is lower in general.

We investigate the reason why the decreases before and after the transit only occur in the mentioned phase angle range.
At $\alpha_\mathrm{p} = 30\degr$, $F_\mathrm{p}P_\mathrm{p}$ is concentrated in two regions which are on the edge of the visible planetary disk but not in the orbital plane and thus outside the areas the moon can occult. As a consequence, a decrease in $P$ at the beginning and end of a transit does not occur. For larger phase angles of more than $120\degr$, the decrease is not observed either. The moon still blocks a high fraction of $F_\mathrm{p}P_\mathrm{p}$ when it crosses the planet's edge. However, this fraction of $F_\mathrm{p}P_\mathrm{p}$ now corresponds to a large fraction of $F$ as well, as the scattered radiation flux of the planet is limited to a relatively small crescent.
$F_\mathrm{p}P_\mathrm{p}$ is even more concentrated on the edge of the planetary disk for wavelengths of $550\ \mathrm{nm}$ and $700\ \mathrm{nm}$ than for $400\ \mathrm{nm}$.
Consequently, a decrease in $P$ before and after the transit is observed over a larger range of phase angles.

We find another change in the behavior when the disk-integrated direction of polarization of the radiation scattered by the planet changes at $\alpha_\mathrm{p} = 166\degr$ for $400\ \mathrm{nm}$ or at $\alpha_\mathrm{p} = 171\degr$ for $550\ \mathrm{nm}$. At a wavelength of $700\ \mathrm{nm}$, the direction of polarization does not change at large phase angles and consequently, the corresponding feature disappears completely.
The degree of linear polarization now decreases by at most $0.75\%$ during a transit. 

As an example, a polarization map at $172\degr$ is shown in the bottom row of Fig.~\ref{fig:transit_polarization}.
$F_\mathrm{p}P_\mathrm{p}$ is largest at the illuminated crescent and the polarization directions are vertical to the planetary edge. In the area, where the moon occults the crescent, $\chi$ is parallel to the orbital plane. For polarization vectors on the crescent in a greater distance of the orbital plane, we always find a vector in a region of approximately equal $F_\mathrm{p}P_\mathrm{p}$ that lies perpendicular to the first one. Therefore, these areas do not contribute to the disk-integrated polarized flux. Around the poles of the planet, $F_\mathrm{p}P_\mathrm{p}$ is too low to contribute significantly. The moon therefore covers the part of the crescent that causes a disk-integrated degree of polarization, so $P$ decreases significantly.

This feature is much smaller at a wavelength of $550\ \mathrm{nm}$ because the zero-crossing of the planetary polarization phase curve occurs at a larger phase angle. As for a wavelength of $700\ \mathrm{nm}$, we instead find a direct transition into the final behavior close to a planetary phase angle of $180\degr$ which happens at $178\degr$ for a wavelength of $400\ \mathrm{nm}$: $P$ is close to 0 again due to the symmetry of the system. During a transit, $P$ now increases again. The moon again breaks the symmetry during the transit.

\subsubsection{Central shadows}
\label{subsubsec:central_shadow}

\begin{figure*}
        \centering
        \includegraphics{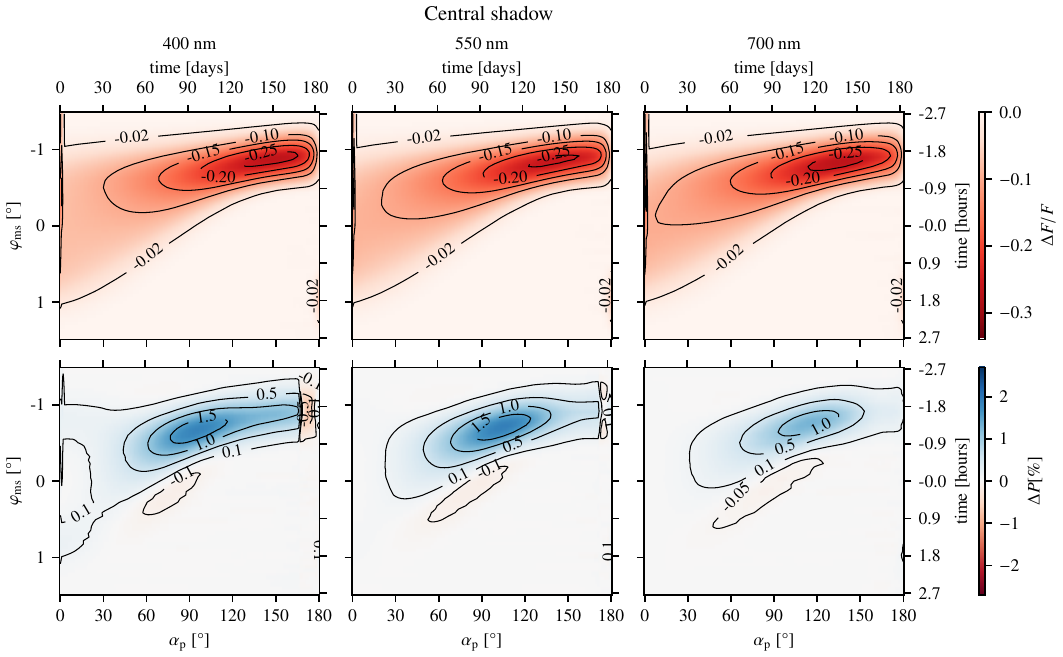}
        \caption{Similar to Fig.~\ref{fig:transit}, but for phases during which the moon casts a shadow on the planet. The left ordinate now shows the angle $\varphi_\mathrm{ms}$, which specifies the lunar position. Again, this angle was transformed to a time in hours, as shown on the right ordinate, where a time of 0 corresponds to the time when the planet, moon, and star are perfectly aligned. A single run of the shadow of the moon across the planetary surface corresponds to a passage through the color maps perpendicular to the abscissa.}
        \label{fig:shadow}
\end{figure*}

The moon in the model system in which all bodies reside within a single plane casts a shadow on the planet if $\varphi_\mathrm{ms} \approx 0$.
The half width of a shadow phase $\Delta\varphi_\mathrm{ms}$ is the maximum $|\varphi_\mathrm{ms}|$ for which part of the planet is within the lunar penumbra. For the model system, we find an upper limit $\Delta\varphi_\mathrm{ms} < 1.49\degr$ (see Appendix~\ref{app:angles}).
A shadow phase in the model system is therefore no longer than $5.4$ hours.
Similar to transits,
the duration of a shadow phase decreases with increasing phase angle \citep[for an analytical approach, see][]{cabrera2007,limbach2024}.

Figure~\ref{fig:shadow} shows the relative flux change $\Delta F/F$ during a shadow phase and the absolute change in the degree of linear polarization $\Delta P$. The color maps have a resolution of $181$ phase angles between $0\degr$ and $180\degr$ 
and a resolution of $64$ lunar position angles $\varphi_\mathrm{ms}$ between $\pm 1.49\degr$. The influence of a single shadow phase is found along an almost vertical line through the diagram. For the model system, a shadow phase occurs 6 to 7 times during half the planetary revolution around the central star.
As expected, the influence of shadows cast on the planet by the moon resembles the influence of lunar transits in front of the planet. We therefore focus on the observed differences.

The maximum values of the relative flux change $\Delta F/F$ during a shadow phase are smaller than for transits at all wavelengths and phase angles. Because the umbra of the moon is small compared to its penumbra, the darkening of a region on the planet by the shadow of its moon causes a smaller $\Delta F$ than the occultation of the same region during a transit \citep{berzosa_molina2018}. 
The largest reductions are $-26.8\%$ found at $146\degr$ for a wavelength of $400\ \mathrm{nm}$, $-25.9\%$ at $141\degr$ for $550\ \mathrm{nm}$ and $-26.9\%$ at $133\degr$ for $700\ \mathrm{nm}$. Again, the influence of the optical depth or the wavelength on $\Delta F/F$ is small.

For $\alpha_\mathrm{p}\leq 2\degr$, the additional influence of a transit is visible again, this time causing a reduction in $\Delta F/F$ at the beginning of the eclipse that corresponds to the end of the transit.
As for transits, the maximum values $\Delta F/F$ reached during a shadow phase increase with increasing phase angle because the illuminated crescent of the planetary surface is smaller at longer wavelengths. Starting at phase angles of $\alpha_\mathrm{p} \approx 140\degr$, $\Delta F/F$ decreases again due to the influence of multiple and forward scattering.
The small flux changes of about $2\%$ to $5\%$ that occur at phase angles larger than $176\degr$ after the shadow phase ended are caused by the shadow falling on the atmosphere on the side of the planet opposite to the crescent.

The changes $\Delta P$ in Fig.~\ref{fig:shadow} show features that are related to the ones found for transits in Fig.~\ref{fig:transit}. The explanation
for transits applies analogously to shadow phases. 
As the decrease in the flux is smaller for shadow phases, the positive changes in the degree of linear polarization are also smaller, because they are mainly the consequence of a decrease in $F$. For a wavelength of $400\ \mathrm{nm}$, the maximum increase in $P$ is $+1.8\%$ for a shadow at a phase angle of $96\degr$. \citet{berzosa_molina2018} find a similar value at $450\ \mathrm{nm}$. At the other wavelengths, we find $+1.8\%$ at $106\degr$ at $550\ \mathrm{nm}$ and $+1.3\%$ at $116\degr$ at $700\ \mathrm{nm}$. As for transits, the phase angle with the maximum increase shifts toward larger phase angles with increasing wavelength.

Contrary to transits, a slight decrease in $P$ is only observed when the shadow is on the illuminated crescent but not at the beginning of a shadow phase. A shadow never falls on the planetary night side, and its shape that appears to an observer is not circular when it first reaches the planetary surface at the terminator. Together, this prevents a decrease in $P$.

Similar to transits, a decrease in $P$ of up to $0.7\%$ is observed at a wavelength of $400\ \mathrm{nm}$ after the direction of polarization of the planet changes. The feature again disappears at a wavelength of $700\ \mathrm{nm}$ and occurs in a very small range of phase angles for a wavelength of $550\ \mathrm{nm}$. These changes appear to be slightly more pronounced for central shadows than for central transits.

\subsubsection{Noncentral transits and shadows}
\label{subsubsec:non_central}

In Sect. \ref{subsubsec:central_transit} and \ref{subsubsec:central_shadow}, we have shown that $\Delta P$ depends on the exact part of the planet that is occulted by the moon or in its shadow. Only if the moon is within the planetary orbital plane, $\Delta P$ takes the values found in the previous sections.

If $i_\mathrm{b} = 90\degr$ and $i_\mathrm{m} \neq 0$, the moon may also transit the planet along a noncentral path and similarly its shadow may take a noncentral path on the planetary surface. We mimic the effect of a nonzero inclination of the lunar orbit by placing the moon in a distance of $r_\mathrm{p}/2$ above the planetary orbital plane in every simulation while keeping its distance to the planet $d_\mathrm{pm}$ constant. In the case of the model system, such a configuration is possible if $i_\mathrm{m} > 0.48\degr$, a small value compared to the Earth's moon with $i_\mathrm{m} \approx 5\degr$.
For the observer, the moon now crosses the planetary disk above the planetary orbital plane, as sketched in Fig.~\ref{fig:transit_polarization}.

To specify the occurrences of transiting and shadowing events, we define $\varphi_\mathrm{ms}^{\prime}$ as the angle between the direction from the center of the planet toward the star and the direction from the center of the planet toward a projection of the center of the moon on to the planetary orbit plane. Now a noncentral transit occurs if $\alpha_\mathrm{p} + \varphi_\mathrm{ms}^{\prime} \approx 360\degr$ and a noncentral shadow falls on the planet if $\varphi_\mathrm{ms}^{\prime} \approx 0$.

\begin{figure}
    \centering
    \includegraphics{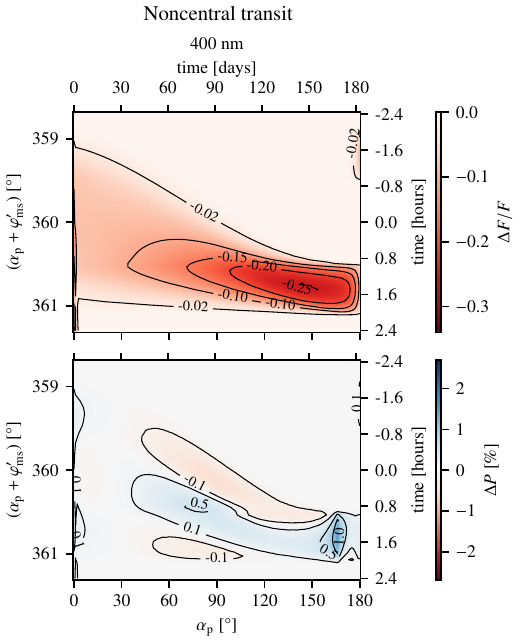}
    \caption{Same as the first column of Fig.~\ref{fig:transit}, but for the results of our simulations with noncentral transits.}
    \label{fig:inclined_transit}
\end{figure}

For a wavelength of $400\ \mathrm{nm}$ only, we repeated the simulations for transits and shadows as described in Sect. \ref{subsubsec:central_transit} and  \ref{subsubsec:central_shadow}, respectively, with the repositioned moon.
Our results for the relative flux change $\Delta F/F$ and the absolute change in polarization $\Delta P$ for noncentral transits are shown in Fig.~\ref{fig:inclined_transit}.

Due to the change in the geometry, noncentral transits last shorter than central ones. The maximum decrease of the flux is slightly smaller than for central transits. Values of $-25.2\%$ are reached for a transit at a phase angle of $141\degr$. However, the shape of the observed $\Delta F/F$ curves remains unchanged compared to central transits.
A moon with a smaller radius on a slightly narrower orbit around the planet can therefore cause the same changes in the observed total flux.

However, the changes $\Delta P$ in the degree of linear polarization behave differently for a noncentral transit, as we see from a comparison of Fig.~\ref{fig:transit} and Fig.~\ref{fig:inclined_transit}.
While for central transits, a maximum increase of $2.6\%$ is found at a phase angle of $98\degr$, the change in the degree of linear polarization is less than $0.5\%$ at the same phase angle for a noncentral transit.
A decrease in $P$ is observed at the end and the beginning of the transit over a larger range of phase angles. Therefore,  $\Delta P$ inhabits a sequence of a decrease, an increase, and a decrease, all with absolute values of $\Delta P$ between $0.1\%$ and $0.5\%$, during noncentral transits at phase angles around $90\degr$. This wave-like behavior of $\Delta P$ is different from the pronounced central peak found for central transits at the same phase angles, where $\Delta P$ reached $+2.5\%$.

For the specific type of noncentral transits described above, the highest increase $\Delta P$ of about $1.5\%$ is observed at a phase angle of $166\degr$. This is the phase angle at which the direction of polarization changes ($P$ is equal to zero for the model system at this phase angle). For central transits, we observe very small changes at $\alpha_\mathrm{p} = 166\degr$. 
Also, a decrease in the degree of polarization is not found at phase angles above $166\degr$. Instead, $P$ increases slightly during a noncentral transit at those phase angles.

The new behavior outlined in this section is explained using Fig.~\ref{fig:transit_polarization} again.
For a noncentral transit at a phase angle of $75\degr$, the moon occults parts of the edge of the visible planetary disk for a longer period of time, which causes the decrease in polarization that is connected to this to last longer as well. Additionally, the moon never covers the region in the orbital plane between the terminator and the edge. The occultation of this region is however connected to the highest increases for central transits around $100\degr$. Therefore, smaller maximum values are observed for noncentral transits at these phase angles.

In the case of a phase angle of $166\degr$, the polarized flux is slightly smaller in the orbital plane than where the moon crosses the planetary edge. The net polarization is zero without the transit because the equivalent region on the lower half of the planet has a similar $F_\mathrm{p}P_\mathrm{p}$ and directions of polarization perpendicular to the ones on the upper half.
However, during the transit, the region on the lower half of the planet contributes to the disk-integrated polarized flux.
Consequently, $P$ increases considerably.
The direction of polarization is diagonal to the orbital plane during such a transit.
For a similar reason, $P$ increases for noncentral transits at higher phase angles. As shown in the panel for $172\degr$ in Fig. \ref{fig:transit_polarization}, the exomoon does not cover the part of the crescent in the orbital plane that causes the disk-integrated degree of polarization if it transits noncentrally at this phase angle. Instead, it breaks the symmetry between the upper and lower half of the visible planetary disk.

\begin{figure}
    \centering
    \includegraphics{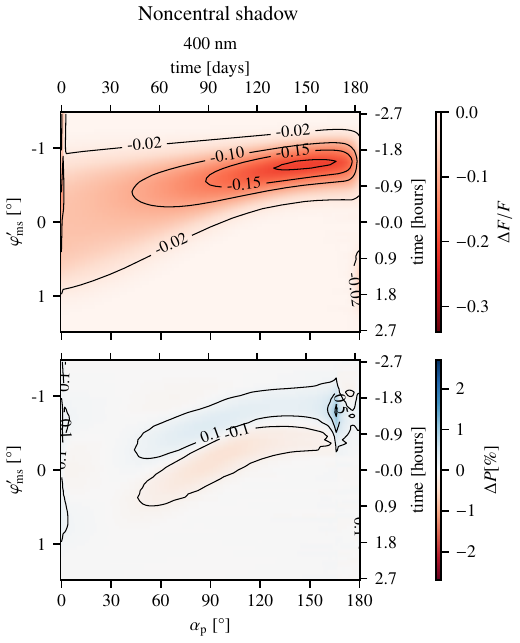}
    \caption{Same as the first column of Fig.~\ref{fig:shadow}, but for the results of our simulations with noncentral shadow phases.}
    \label{fig:inclined_shadow}
\end{figure}

Similar to central transits and shadows, noncentral transits and shadows also cause connatural results, as is clear from our results for noncentral shadows shown in Fig.~\ref{fig:inclined_shadow}.
The decrease in the degree of linear polarization at the beginning of a noncentral shadow phase is again much smaller than for a transit because the shadow never falls on the night side of the planet and thus never crosses the terminator.
The decrease at the end of a shadow phase is observed for a larger range of phase angles. For a phase angle around $90\degr$, the resulting changes in $\Delta P$ are therefore an increase during the first half of the event followed by a decrease in the second half, with similar maximum absolute values of $\Delta P$ during both halves.

Clearly, the behavior of $\Delta P$ depends on the exact path of the moon or its shadow across the visible planetary disk. Mutual events caused by a moon can thus show large variations in their influence on the linear polarization, even when two events in the same system are compared.

\section{Discussion}
\label{sec:discussion}

\subsection{Resolved and unresolved observability}

So far, we have only considered the flux of the planet and the moon but not the stellar flux. This resembles observations during which the planet and the star are spatially resolved. Such observations of the scattered radiation of exoplanets are rarely possible today. Until now, directly imaged planets are either massive gas giants in large separations from the host star \citep[e.g., $\beta$ Pic b,][]{lagrange2009} or accreting planets still within a protoplanetary disk that were discovered through detections of their thermal emission \citep[e.g., PDS 70 b,][]{keppler2018}. Massive gas giants in large separations from the host star are likely a promising target to discover exomoons \citep[see also][]{lazzoni2022}. However, high-precision polarimetry is only promising for close-in exoplanets at the moment.

Observing the scattered radiation of Earth-like exoplanets directly is a target of the Habitable Worlds Observatory, a future space telescope in the 2040s, which will be equipped with a state-of-the-art coronagraph \citep{NAP2023,vaughan2023,tuchow2024}. 
The HWO will be able to detect mutual events in systems analogous to the Earth and its moon up to a distance of $10\ \mathrm{pc}$ through the variation in flux \citep{limbach2024}.
Provided that polarimetric observations with the required precision of about $0.1\%$ to $1\%$ are also possible, the influence of an exomoon on the linear polarization of its exoplanet will potentially be detectable with the HWO.

On the other hand, if the stellar flux is included in the calculations, the degree of polarization caused by the planet and the moon together is less than $2\cdot 10^{-10}$ at all wavelengths of the considered model system. This is already clearly below the accuracy of a few parts per million reached by modern polarimeters \citep{bailey2020,wiktorowicz2023}. The influence of mutual events on the linear polarization is therefore on the order of $10^{-12}$. Even the intrinsic polarization of radiation emitted by inactive stars of about $1$ to $10\ \mathrm{ppm}$ \citep{kemp1987,cotton2017} and its variability have to be considered very carefully to evaluate the possible impact of exomoons.
While a particularly bright Hot Jupiter might cause a polarization of over $10\ \mathrm{ppm}$ \citep[e.g.,][]{bailey2021}, 
it is unclear whether Hot Jupiters can even host exomoons, as they might lose their exomoons during their migration to an orbit close to the host star \citep{trani2020}.

\subsection{Usability of polarimetry for exomoon detections}
\label{subsec:usability}

To detect unknown exomoons through the observation of mutual events, the system has to be observed with a high duty cycle to prevent missing mutual events. For the model system, a mutual event only lasts between $1$ and $5$ hours. Therefore, such a system has to be observed with a time resolution of about $0.5$ hours for the time of one lunar orbit period, that is about a month. Such observations are only possible from space \citep[see][]{cabrera2007}. In a system with higher inclinations, mutual events occur less often, which demands observations over an even longer time period.

In systems with smaller lunar orbital radii than the Earth-Moon system, mutual events occur more frequently and such systems are consequently easier to detect \citep[see also][]{cabrera2007,limbach2024}.
We note that while the time axis of our plots would have to be adapted accordingly, the changes $\Delta F/F$ and $\Delta P$ during a transit as a function of $\varphi_\mathrm{ms}$ are not affected by a smaller lunar orbital radius if the mutual illumination of the two bodies remains negligible except for combined transit and shadowing events at small phase angles. In the case of shadows, the values found for $\Delta F/F$ and $\Delta P$ would not be valid in the case of smaller lunar orbital radii.

Because the change in the polarization $\Delta P$ is highly dependent on the lunar orbit, the flux change $\Delta F$ should be measured simultaneously. To confirm that a measured change $\Delta F$ is caused by a moon, a prediction for future mutual events has to be made.
The measurements of $\Delta P$ are then useful to narrow down the parameter space of the lunar orbital parameters. In Sect. \ref{subsubsec:non_central}, we showed that the path of the moon or its shadow over the planetary disk has a significant influence on the results and can be used to set limits on the inclination of the lunar orbit.
After possible paths are extracted from measurements of $\Delta P$, the radius of the moon and its orbital speed can be estimated from the duration of the mutual event. These constraints are sufficient to predict possible periods of time in which the next mutual event occurs. To confirm the existence of the exomoon, the system should then be observed during this period.

In some cases, it might also be possible to verify the presence of an exomoon with different approaches, for instance through transit timing or transit duration variations \citep[e.g.,][]{simon2007,kipping2009} or the orbital sampling effect \citep{heller2014} in systems where the exoplanet transits the star or through observations at wavelengths where the moon outshines the planet \citep[e.g., in the $1.4\ \text{µm}$ water absorption band, see][]{limbach2024} using spectroastrometric observations \citep{agol2015}.

This work does not take into consideration that the geometrical albedo of Earth-like exoplanets might vary significantly.
An inhomogeneous surface, such as continents and oceans on Earth, can lead to a short-term variation of the polarized and total scattered radiation flux due to the planetary rotation \citep{song2017,sterzik2019,wang2019,groot2020}.
Such short periodic variations should not be mistaken for a moon. Variations in the planetary phase curves generally make it harder to detect an exomoon \citep[see][]{limbach2024}. Consequently, exo\-planets with a less variable polarization phase curve offer a better chance to detect an exomoon. Planets covered in a dense cloud layer are therefore a reasonable expansion of our study.

\section{Conclusion}
\label{sec:conclusion}

We presented simulation results for the influence of an exomoon similar to the Earth's moon on the flux and polarization phase curves of an Earth-like exoplanet at wavelengths of $400\ \mathrm{nm}$, $550\ \mathrm{nm}$, and $700\ \mathrm{nm}$. We found that an inclined orbit of the exomoon causes considerable variations in the polarization signal observed during mutual events.

In Sect. \ref{subsec:phase curve} we first calculated the flux and linear polarization phase curves of the planet ($F_\mathrm{p}(\alpha), P_\mathrm{p}(\alpha)$) and the moon ($F_\mathrm{m}(\alpha), P_\mathrm{m}(\alpha)$) of the model presented in Sect. \ref{sec:model} and of the planet and moon combined ($F(\alpha)$, $P(\alpha)$) using the three-dimensional Monte Carlo radiative transfer code POLARIS \citep{reissl2016,lietzow2021}. For $F(\alpha)$ and $P(\alpha)$, we assumed that the planet-moon system is observed spatially resolved from the host star, while the planet and the moon remain spatially unresolved. The moon is assumed to be a diffuse reflecting sphere in the model ($P_\mathrm{m} = 0$).
Neglecting the influence of mutual events, the degree of linear polarization of the planet alone is therefore slightly higher compared to the two bodies combined, with a difference of less than $1\%$ at all wavelengths, which is similar to the results by \citet{berzosa_molina2018}.
However, we showed that these findings cannot be generalized. If the orientation of polarization of the disk-integrated planetary and lunar scattered radiation were identical and $P_\mathrm{m}$ were greater than $P_\mathrm{p}$, the degree of linear polarization would be higher for the overall system of an exoplanet and its moon combined than for the planet alone.
As pointed out by \citet{berzosa_molina2018}, an ambiguity remains even if the small differences could be measured. A slightly adapted planetary model can also explain a slightly higher flux and slightly changed degree of linear polarization.
This ambiguity can be resolved through observations of mutual events. 
Our results for the influence of mutual events were presented in Sect. \ref{subsec:mutual_events}.

For occultations of the moon by the planet or lunar eclipses during which the moon is in the shadow of the planet, we calculated the changes in the total flux $\Delta F$ and linear polarization $\Delta P$ analytically in Sect. \ref{subsubsec:occultation_eclipse}. For the model system, the biggest change in the linear polarization is an increase of $\Delta P = 0.77\%$ for a lunar darkening event at $\alpha = 91\degr$ at a wavelength of $400\ \mathrm{nm}$ and $+0.46\%$ for a lunar darkening event at $550\ \mathrm{nm}$. This is less than the values of around $2.5\%$ which were previously reported by \citet{berzosa_molina2018} for a similar system, although their flux values suggest a $\Delta P$ comparable to ours. 
Again, we found that for identical directions of polarization and $P_\mathrm{m} > P_\mathrm{p}$, the degree of linear polarization instead decreases during these mutual events.
We discussed that with observations of these types of mutual events, it is possible to disentangle the linear polarization phase curves of the exoplanet and the exomoon.

The results from our polarized radiative transfer simulations of transits of the moon in front of the planet and time periods during which the moon casts a shadow on the planet were discussed in detail in Sect. \ref{subsubsec:central_transit} to \ref{subsubsec:non_central}. By using a Monte Carlo radiative transfer code, we were able to calculate the influence even for larger phase angles for which a completely spherical description of the planetary atmosphere is necessary. In Sect. \ref{subsubsec:non_central}, we placed the moon at a distance of $r_\mathrm{p}/2$ to the planetary orbital plane to mimic the influence of an inclined exomoon orbit. This resulted in noncentral transits and shadows.

In general, the influence of a transit is higher than the influence of a shadow because of the relatively large penumbra compared to the umbra in the model system \citep{berzosa_molina2018}.
The highest relative flux changes $\Delta F/F$ are observed at large phase angles $\alpha_\mathrm{p}$ when the moon is in the planetary orbital plane. At these phase angles, only a small crescent of the illuminated day side of the planet is visible to the observer. The moon occults a large portion of this crescent or darkens a large portion with its shadow. For transits, $\Delta F/F$ decreases at most by $32\%$ to $36\%$. For shadows, the changes are smaller with decreases of $25\%$ to $27\%$.
For noncentral mutual events during which the moon has a distance of $r_\mathrm{p}/2$ to the orbital plane of the planet, $\Delta F/F$ is smaller, reaching values of $25\%$ for transits and $20\%$ for shadows.
From decreases in the flux alone it is hard to narrow down the space of possible exomoon properties because a smaller moon on an orbit with a smaller semimajor axis causes a very similar signal to the noncentral events. 

We used radiative transfer simulations to also calculate the changes in the degree of linear polarization $\Delta P$ during a transit. For wavelengths of $400\ \mathrm{nm}$ and $550\ \mathrm{nm}$ the absolute value of $\Delta P$ is slightly higher than for $700\ \mathrm{nm}$.
Similar to \citet{berzosa_molina2018}, our simulations showed that increases are expected for central transits and shadows. However, we found that at large phase angles where the direction of polarization of the planet changes the degree of polarization decreases instead.
The maximum increase in $P$ is about $2.7\%$ for wavelengths of $400\ \mathrm{nm}$ and $550\ \mathrm{nm}$ for central transits and about $1.8\%$ for central shadows. For $700\ \mathrm{nm}$, we find smaller increases of $1.9\%$ for transits and $1.3\%$ for shadows. 
Identical to the results by \citet{berzosa_molina2018} for $450\ \mathrm{nm}$, these maxima occur for transits or shadows observed at a phase angle of $96\degr$ to $106\degr$ just after quadrature, which  offer the highest angular distance between the planet and the star.
For noncentral transits and shadows, only a small increase of less than $0.5\%$ is found at a wavelength of $400\ \mathrm{nm}$ in this phase angle range. Additionally, $\Delta P$ inhabits a wave-like sequence of increases and decreases that are similar in their absolute value instead of a dominant central peak. The largest increase in linear polarization during a noncentral mutual event is instead found at a phase angle of $166\degr$, where the direction of polarization changes and is 0 without the mutual event. For noncentral transits $\Delta P = 1.5\%$ at this phase angle, for noncentral shadows $\Delta P = 1.1\%$.

Because the maximum values and the shape of the linear polarization signal are dependent on the exact path that the moon or its shadow takes over the visible planetary disk seen by the observer, it can be used to narrow down possible values for the inclination of the lunar orbit. Therefore, measurements of $\Delta P$ during a mutual event facilitate the prediction of the timing of future mutual events, as discussed in Sect. \ref{subsec:usability}.

If the planet-moon system were spatially unresolved from the host star, the stellar flux would dominate the total flux. Consequently, the resulting linear polarizations would be on the order of $10^{-12}$. This is far below the precision of a few parts per million  reached with modern polarimeters \citep{bailey2020,wiktorowicz2023}. However, enhanced polarimeters or co\-ro\-na\-gra\-phic observations 
might be able to measure the influence of mutual events of Earth-Moon analogs.

\begin{acknowledgements}
This research made use of NASA’s Astrophysics Data System (\url{https://ui.adsabs.harvard.edu}), Astropy, a community-developed core Python package for Astronomy \citep{astropy2013,astropy2018,astropy2022}, Matplotlib (\url{https://matplotlib.org/}) \citep{hunter2007} and Numpy (\url{https://numpy.org/}) \citep{harris2020}. It was supported in part through high-performance computing resources available at the Kiel University Computing Center. M.M. and S.W. thank the DFG for financial support under grant WO 857/24-1. The authors thank the anonymous referee for constructive comments and suggestions.
\end{acknowledgements}

\bibliography{references.bib}

\begin{appendix}

\section{Moon-to-planet flux ratios}
\label{app:flux_ratios}

\begin{figure}[hp!]
        \includegraphics{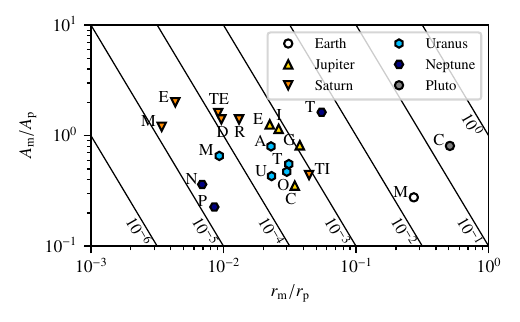}
        \caption{Ratios of the lunar flux to the planetary flux $F_\mathrm{m}/F_\mathrm{p}$  computed according to Eq. \eqref{eq:F_ratio}. $F_\mathrm{m}/F_\mathrm{p}$ depends on the ratio of the lunar to the planetary radius $r_\mathrm{m}/r_\mathrm{p}$ and the ratio of the geometrical albedoes $A_\mathrm{m}/A_\mathrm{p}$. Contour lines indicate equal flux ratios. Symbols indicate values for planet-moon pairs in the Solar System at a wavelength of $549\,\mathrm{nm}$. The different planets are marked by different symbols; moons are indicated by their first letter(s). Circumplanetary rings are not included in the calculation. All values are from the NASA planetary fact sheets (D.~R. Williams, \url{https://nssdc.gsfc.nasa.gov/planetary/planetfact.html}, as of May 7, 2024). \\
\emph{Abbreviations of the moons:} Earth: M (The Moon); Jupiter: I (Io), E (Europa), G (Ganymede), C (Callisto); Saturn: M (Mimas), E (Enceladus), TE (Tethys), D (Dione), R (Rhea), TI (Titan); Uranus: Ariel (A), Umbriel (U), Titania (T), Oberon (O), Miranda (M); Neptune: Triton (T), Nereid (N), Proteus (P); Pluto: Charon (C).}
        \label{fig:app_basic_flux}
\end{figure}

We assume that a planet and a moon are illuminated only by their central star and neglect the mutual illumination of the planet and the moon. The ratio of the flux reflected by a moon to the flux reflected by its planet is now approximated using their geometrical albedoes $A_\mathrm{m}, A_\mathrm{p}$ and radii $r_\mathrm{m}, r_\mathrm{p}$,
\begin{equation}
        \frac{F_\mathrm{m}}{F_\mathrm{p}} \approx \frac{A_\mathrm{m}}{A_\mathrm{p}}\frac{r_\mathrm{m}^2}{r_\mathrm{p}^2}.
        \label{eq:F_ratio}
\end{equation}
Results from this equation for the moons in the Solar System at a wavelength of $550\ \mathrm{nm}$ are shown in Fig.~\ref{fig:app_basic_flux}. The approximation is exact for a phase angle of $0^\circ$. Be aware that it does not include an additional influence of circumplanetary rings.

\section{Mutual illumination}
\label{app:mutual_illumination}

While a single planet is only illuminated by its host star, a celestial body in a planet-moon system is also illuminated by the other body in the system.
The overall flux of the system is made up of different photon paths which we group into different orders.

The zeroth-order flux is the flux of stellar radiation reaching the observer without interacting with the planet or the moon. This flux is \citep[e.g.,][]{zalewski1995,stam2004}
\begin{equation}
        F_\mathrm{s} = \frac{\pi r_\mathrm{s}^2}{d_\mathrm{so}^2}B_\nu(T),
\end{equation}
where $d_\mathrm{so}$ is the distance between the star and the observer.

We now define the first-order flux to be the flux of stellar radiation reflected toward the observer by either the planet ($F_\mathrm{1p}$) or the moon ($F_\mathrm{1m}$) without interacting with the other celestial body. The first-order flux is therefore the total flux neglecting the mutual illumination if the planet and the moon are unresolved but the planet-moon system is resolved from the star. The fluxes $F_\mathrm{1p}$ and $F_\mathrm{1m}$ depend on the phase angle $\alpha_\mathrm{p}$ of the planet and $\alpha_\mathrm{m}$ of the moon. The flux phase curves of the two celestial bodies are thus
\begin{equation}
        F_\mathrm{1p}(\alpha_\mathrm{p}) = F_\mathrm{p0} A_\mathrm{p} \psi_\mathrm{p}(\alpha_\mathrm{p}),
        \quad
        F_\mathrm{1m}(\alpha_\mathrm{m}) = F_\mathrm{m0} A_\mathrm{m} \psi_\mathrm{m}(\alpha_\mathrm{m}).
        \label{eq:first_order}
\end{equation}
Here, $F_\mathrm{0p}$ is defined in Eq. \eqref{eq:F0p} and $F_\mathrm{0m}$ is defined similarly. $A_\mathrm{p}$ and $A_\mathrm{m}$ are the geometric albedoes.
Also, we use the phase law $\psi_\mathrm{p}(\alpha) = F_\mathrm{1p}(\alpha)/F_\mathrm{1p}(0)$ of the planet and the phase law $\psi_\mathrm{m}$ of the moon that is defined analogously \citep[see][]{lester1979}.

Flux of the $n$-th order is scattered by either the planet or the moon $n$ times. Therefore, second-order flux is the flux that was reflected by the planet or the moon toward the other celestial body once before being scattered toward the observer. If $r_\mathrm{m} \ll d_\mathrm{pm}$, the flux of the stellar radiation reflected by the moon measured on the planetary surface is
\begin{equation}
        F_\mathrm{1m,p} = \pi B_\nu(T) \frac{r_\mathrm{s}^2 r_\mathrm{m}^2}{d_\mathrm{sm}^2 d_\mathrm{pm}^2} A_\mathrm{m} \psi_\mathrm{m}(\pi - \varphi_\mathrm{ms}),
\end{equation}
where $\varphi_\mathrm{ms}$ is defined in Fig.~\ref{fig:geometry}. The second-order flux of the planet $F_\mathrm{2p}$ is therefore calculated by replacing the stellar flux on the planetary surface $\pi B_\nu(T) r_\mathrm{s}^2/d_\mathrm{sp}^2$ in Eq. \eqref{eq:first_order} with the flux $F_\mathrm{1m,p}$. Consequently, using $d_\mathrm{po}\approx d_\mathrm{mo}$
\begin{align}
        F_\mathrm{2p}(\varphi_\mathrm{ms},\varphi_\mathrm{mo}) & = \pi B_\nu(T) \frac{r_\mathrm{s}^2 r_\mathrm{p}^2 r_\mathrm{m}^2}{d_\mathrm{sm}^2 d_\mathrm{pm}^2 d_\mathrm{po}^2} A_\mathrm{p} A_\mathrm{m} \psi_\mathrm{m}(\pi - \varphi_\mathrm{ms})\,\psi_\mathrm{p}(\varphi_\mathrm{mo}) \notag \\
        & = F_\mathrm{1m,max} \frac{r_\mathrm{p}^2}{d_\mathrm{pm}^2} A_\mathrm{p} \psi_\mathrm{m}(\pi - \varphi_\mathrm{ms})\,\psi_\mathrm{p}(\varphi_\mathrm{mo}).
\end{align}
where $\varphi_\mathrm{mo}$ is the angle between the direction toward the moon and the observer at the center of the planet. For a non-inclined system as in Fig.~\ref{fig:geometry}, we have $\varphi_\mathrm{mo} = \varphi_\mathrm{ms} + \alpha_\mathrm{p}$.
$F_\mathrm{1p,max}$ and $F_\mathrm{1m,max}$ are the maximum values of the first-order fluxes.
For the second-order flux of the moon $F_\mathrm{2m}$, we get
\begin{equation}
        F_\mathrm{2m}(\varphi_\mathrm{ms},\varphi_\mathrm{mo}) = F_\mathrm{1m,max} \frac{r_\mathrm{p}^2}{d_\mathrm{pm}^2} A_\mathrm{p} \psi_\mathrm{m}(\varphi_\mathrm{ms})\,\psi_\mathrm{p}(\pi-\varphi_\mathrm{mo})
\end{equation}
as the distances of the planet or the moon to the central star are comparable, $d_\mathrm{sp} \approx d_\mathrm{sm}$.
Now, we set a boundary for the second-order flux maximum in comparison to the first-order influence of the moon,
\begin{equation}
        \frac{F_\mathrm{2p,max} + F_\mathrm{2m,max}}{F_\mathrm{1m,max}} <
        2\frac{r_\mathrm{p}^2}{d_\mathrm{pm}^2}A_\mathrm{p}.
\end{equation}
For the model system described in Sect. \ref{sec:model}, the right-hand side is less than $1.9\cdot 10^{-4}$. In that case, the second-order flux and thus the mutual illumination are negligible. However, this cannot be generalized to all planet-moon systems. If the moon orbits its planet near the Roche limit at about $2.5r_\mathrm{p}$ \citep[e.g.,][]{williams2003}, the value of the right-hand side rises to about $0.1$. 
Whether the mutual illumination can be neglected when calculating the influence of an exomoon on the phase curve of an exoplanet has to be decided for each system individually.

\section{Half width of shadows and eclipses}
\label{app:angles}

\begin{figure}[!ht]
        \centering
        \includegraphics{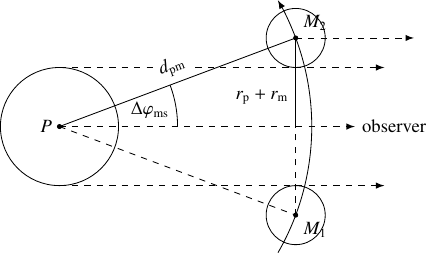}
        \caption{Half width of a transit $\Delta\varphi_\mathrm{ms}$ calculated from the triangle between the direction toward the observer, the center of the planet $P$ and the positions $M_1$ and $M_2$ of the center of the moon at the beginning and the end of the transit. All arrows point toward the observer and are roughly parallel. The figure  is not to scale.}
        \label{fig:transit_width}
\end{figure}

For a transit, the half width is calculated from Fig.~\ref{fig:transit_width},
\begin{equation}
        \Delta\varphi_\mathrm{ms} = \arcsin\frac{r_\mathrm{p} + r_\mathrm{m}}{d_\mathrm{pm}}.
\end{equation}
Here, we assumed that the distance between the observer and the planet is much larger than the distance between the planet and the moon. This is the case for Solar System-based observations of exoplanets and exomoons.

\begin{figure}[!tp]
    \centering
    \includegraphics[width = \linewidth]{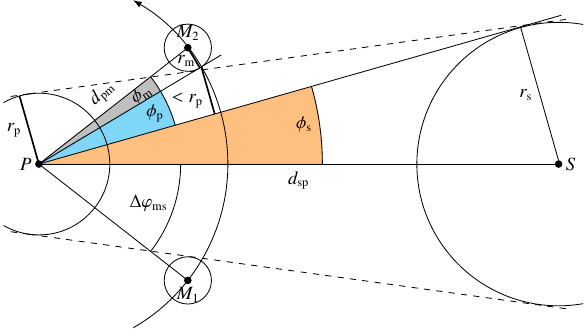}
    \caption{Half width of a shadow phase $\Delta\varphi_\mathrm{ms}$  estimated from the three angles $\phi_\mathrm{m}, \phi_\mathrm{p}$, and $\phi_\mathrm{s}$.}
    \label{fig:shadow_angle}
\end{figure}

The half width of a shadow phase is calculated according to Fig.~\ref{fig:shadow_angle}. In this case, three contributions were identified. The quantity  $\phi_\mathrm{s}$ is the angle under which the star is seen from the planetary center. It is interpreted as describing the time the center of the moon needs to cross the stellar disk from the perspective of an observer on the planet. However, an extended moon needs slightly longer to cross the stellar disk. This effect is included by adding the angle under which the moon is seen from the planetary center, $\phi_\mathrm{m}$. Finally, $\phi_\mathrm{p}$ is added to accommodate the time the shadow of the moon needs to cross the planet's surface. From Fig.~\ref{fig:shadow_angle}, we get
\begin{equation}
    \Delta\varphi_\mathrm{ms} < \arcsin\frac{r_\mathrm{s}}{d_\mathrm{sp}} + \arcsin\frac{r_\mathrm{p}}{d_\mathrm{pm}} + \arcsin\frac{r_\mathrm{m}}{d_\mathrm{pm}},
\end{equation}
where we used an upper limit for $\phi_\mathrm{p}$ that is reached when $d_\mathrm{sp}$ approaches infinity (for a very high distance between the planet and the star).

\section{Influence of the lunar polarization phase curve during transits}
\label{app:lunar_influence}

Here, we estimate the influence of the actual lunar polarization phase curve on the changes $\Delta P$ in the linear polarization that are observed during transits. Again, $F_\mathrm{p}, P_\mathrm{p}$ are the flux and degree of linear polarization of the planet, respectively, and $F_\mathrm{m}, P_\mathrm{m}$ the same quantities for the moon. We assume that the polarized flux of the planet is reduced from $F_\mathrm{p}P_\mathrm{p}$ to $aF_\mathrm{p}P_\mathrm{p}$ and the total flux is reduced from $F_\mathrm{p}$ to $bF_\mathrm{p}$, where $a,b \in [0,1]$.
Furthermore, we assume that the direction of polarization is identical for the planet and moon. Assuming parameters similar to the moon of Earth, this is the case at all phase angles between the inversion phase angle of the moon of about $30\degr$ \citep[e.g.,][]{lyot1929,coyne1970,shkuratov1992a} and the change in the direction of polarization at about $166\degr$ for a wavelength of $400\ \mathrm{nm}$. 
Also, the direction of polarization is assumed to be constant across the planetary disk which is approximately true for the above-mentioned range of phase angles (see Fig.~\ref{fig:transit_polarization}).
This results in a change in the degree of linear polarization of
\begin{equation}
    \Delta P = \frac{aF_\mathrm{p}P_\mathrm{p} + F_\mathrm{m}P_\mathrm{m}}{bF_\mathrm{p}+F_\mathrm{m}} - \frac{F_\mathrm{p}P_\mathrm{p} + F_\mathrm{m}P_\mathrm{m}}{F_\mathrm{p}+F_\mathrm{m}},
\end{equation}
thus,
\begin{align}
    \Delta P & = \left(\frac{a}{bF_\mathrm{p} + F_\mathrm{m}} - \frac{1}{F_\mathrm{p} + F_\mathrm{m}}\right) F_\mathrm{p}P_\mathrm{p} \notag \\
    & \hspace{0.5cm} + \left(\frac{1}{bF_\mathrm{p} + F_\mathrm{m}} - \frac{1}{F_\mathrm{p} + F_\mathrm{m}}\right) F_\mathrm{m}P_\mathrm{m}.
    \label{eq:dP_analytic}
\end{align}
If we let $P_\mathrm{m} = 0$, the systematic error in $\Delta P$ is given by the second term of Eq. \eqref{eq:dP_analytic}. If $F_\mathrm{p}/F_\mathrm{m} \gg 1/b$ and $F_\mathrm{p}/F_\mathrm{m} \gg 1$, the term simplifies to
\begin{equation}
    \Delta P_\mathrm{sys} \approx \left(\frac{1}{b}-1\right) \frac{F_\mathrm{m}}{F_\mathrm{p}}P_\mathrm{m}.
    \label{eq:dP_easy}
\end{equation}
In our simulations, $F_\mathrm{p}/F_\mathrm{m} > 43$ at $700\ \mathrm{nm}$ and even higher at wavelengths of $400\ \mathrm{nm}$ and $550\ \mathrm{nm}$. With $1/b < 1.5$, the assumption is justified. We note that the simplification might not be possible if the parameters of the planet and the moon are different, especially when the moon has a similar or higher surface albedo than the planet. In that case, the second term of Eq. \eqref{eq:dP_analytic} should be used instead of Eq. \eqref{eq:dP_easy}.

$\Delta P_\mathrm{sys}$ is always positive for equal polarization directions of the planet and the moon. In our study, using a realistic maximum value of $P_\mathrm{m} = 10\%$ \citep[see][]{lyot1929,coyne1970} leads to $\Delta P_\mathrm{sys} \leq 0.08\%$ for a wavelength of $400\ \mathrm{nm}$.
While this may result in a slight underestimation of positive changes $\Delta P$ and a slight overestimation of negative changes $\Delta P$, it does not change the general behavior shown in Fig.~\ref{fig:transit}. Even the slight decreases in the polarization direction at the end and beginning of a transit are unaffected because $1 - b$ is small when the moon is only partially in front of the edge of the planet or the terminator, so $1/b$ is close to $1$.

\section{Disk resolved polarized flux}
\label{app:disk_resolved_F_P}

\begin{figure*}
    \centering
    \includegraphics{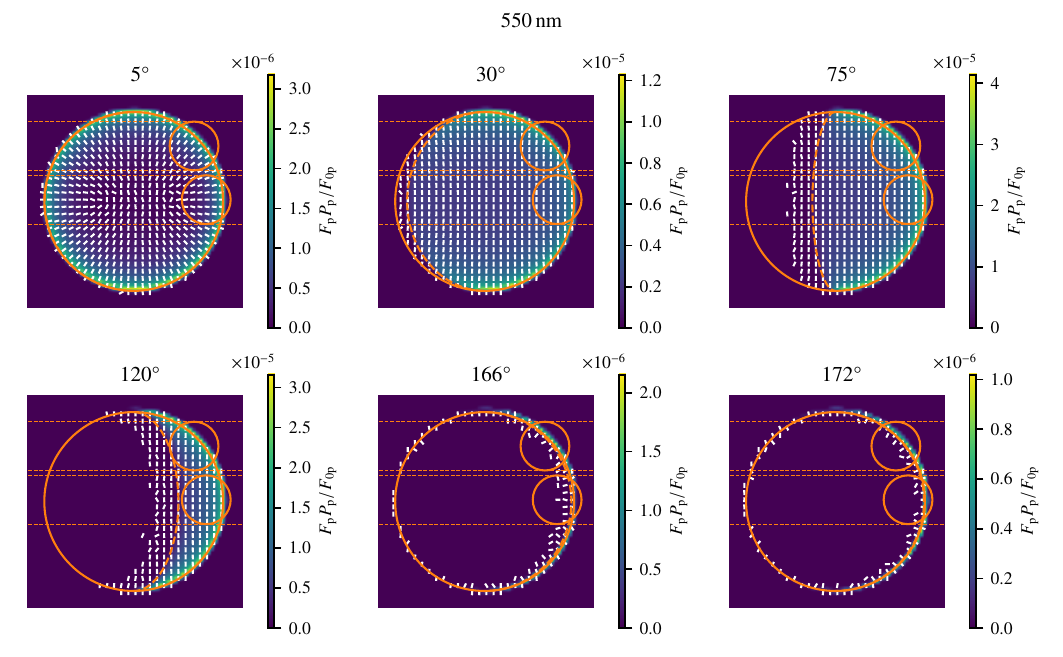}
    \caption{Same as Fig. \ref{fig:transit_polarization}, but for a wavelength of $550\ \mathrm{nm}$.}
    \label{fig:transit_polarization_550}
\end{figure*}
\begin{figure*}
    \centering
    \includegraphics{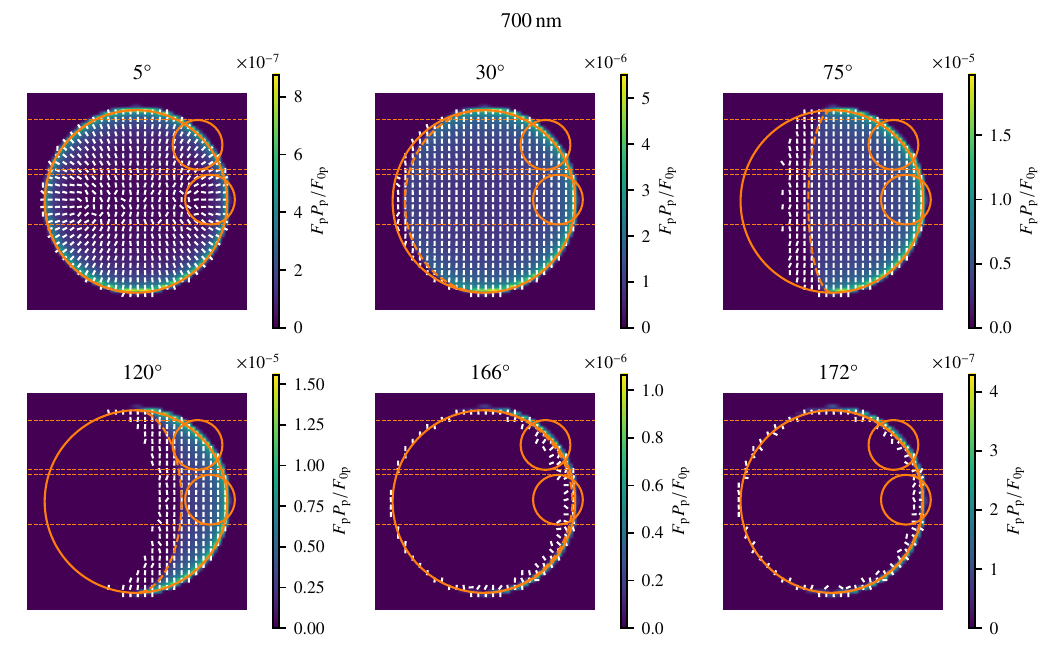}
    \caption{Same as Fig. \ref{fig:transit_polarization}, but for a wavelength of $700\ \mathrm{nm}$.}
    \label{fig:transit_polarization_700}
\end{figure*}

In Fig. \ref{fig:transit_polarization_550} and \ref{fig:transit_polarization_700}, we present results for the disk-resolved polarized flux of the planet in our model system (see Sect. \ref{sec:model}) at wavelengths of $550\ \mathrm{nm}$ and $700\ \mathrm{nm}$, respectively.

\end{appendix}

\end{document}